\documentclass[twocolumn,showpacs,amsmath,floatfix,prb,aps]{revtex4}

\usepackage{color}
\usepackage{amsmath}
\usepackage{amssymb}  
\usepackage{bbold}
\usepackage{float}
\usepackage{pifont}   
\usepackage{graphicx} 
\usepackage{dcolumn}  
\usepackage{bm}       
\usepackage{amsfonts} 
\usepackage{multirow} 

\begin{document}

\newcommand{\ba}{{\bf a}}
\newcommand{\BB}{{\bf b}}
\newcommand{\bd}{{\bf d}}
\newcommand{\br}{{\bf r}}
\newcommand{\bp}{{\bf p}}
\newcommand{\bk}{{\bf k}}
\newcommand{\bg}{{\bf g}}
\newcommand{\bt}{{\bf t}}
\newcommand{\bu}{{\bf u}}
\newcommand{\bq}{{\bf q}}
\newcommand{\bG}{{\bf G}}
\newcommand{\bP}{{\bf P}}
\newcommand{\bJ}{{\bf J}}
\newcommand{\bK}{{\bf K}}
\newcommand{\bL}{{\bf L}}
\newcommand{\bR}{{\bf R}}
\newcommand{\bS}{{\bf S}}
\newcommand{\bT}{{\bf T}}
\newcommand{\bQ}{{\bf Q}}
\newcommand{\bA}{{\bf A}}
\newcommand{\bH}{{\bf H}}

\newcommand{\bra}[1]{\left\langle #1 \right |}
\newcommand{\ket}[1]{\left| #1 \right\rangle}
\newcommand{\braket}[2]{\left\langle #1 | #2 \right\rangle}
\newcommand{\mel}[3]{\left\langle #1 \left| #2 \right| #3 \right\rangle}
\newcommand{\vect}[1]{\boldsymbol{#1}}

\newcommand{\bdel}{\boldsymbol{\delta}}
\newcommand{\bsig}{\boldsymbol{\sigma}}
\newcommand{\beps}{\boldsymbol{\epsilon}}
\newcommand{\bnu}{\boldsymbol{\nu}}
\newcommand{\bnab}{\boldsymbol{\nabla}}

\newcommand{\bgt}{\tilde{\bf g}}

\newcommand{\brh}{\hat{\bf r}}
\newcommand{\bph}{\hat{\bf p}}

\author{M. Vogl$^1$}
\author{O. Pankratov$^1$}
\author{S. Shallcross$^1$}
\affiliation{1 Lehrstuhl f\"ur Theoretische Festk\"orperphysik, Staudtstr. 7-B2, 91058 Erlangen, Germany.}

\title{Semiclassics for matrix Hamiltonians: The Gutzwiller trace formula and applications to the graphene-type systems}
\date{\today}

\begin{abstract}

We have extended the semi-classical theory to include a general account of matrix valued Hamiltonians, i.e. those that describe quantum systems with internal degrees of freedoms, based on a generalization of the Gutzwiller trace formula for a $n\times n$ dimensional Hamiltonian $H(\hat{\bp},\hat{\bq})$. The classical dynamics is governed by $n$ Hamilton-Jacobi (HJ) equations, that act in a phase space endowed with a classical Berry curvature encoding anholonomy in the parallel transport of the eigenvectors of $H(\hat{\bp}\to\bp,\hat{\bq}\to\bq)$; which describe the internal structure of the semi-classical particles. This Berry curvature is a fully classical object and is, in that sense, as fundamental to the semi-classical theory of matrix Hamiltonians as the Hamilton-Jacobi equations. At the $\mathcal{O}(\hbar^1)$ level, it results in an additional semi-classical phase composed of (i) a Berry phase and (ii) a dynamical phase resulting from the classical particles ``moving through the Berry curvature''. We show that the dynamical part of this semi-classical phase will, generally, only be zero only for the case in which the Berry phase is topological (i.e. depends only on the winding number). We illustrate the method by calculating the Landau spectrum for monolayer graphene, the four-band model of AB bilayer graphene, and for a more complicated matrix Hamiltonian describing the silicene band structure. Finally we apply our method to an inhomogeneous system consisting of a strain engineered one dimensional moir\'e in bilayer graphene, finding localized states near the Dirac point that arise from electron trapping in a semi-classical moir\'e potential. The semi-classical density of states of these localized states we show to be in perfect agreement with an exact quantum mechanical calculation of the density of states.
\end{abstract}


\maketitle


\section{Introduction}

In the $\hbar\to 0$ limit the behaviour of quantum systems can be expressed in terms of the classical phase space trajectories. For one dimensional scalar problems the WKB method \cite{Wentzel,Kramers,Brillouin} yields semi-classical wavefunctions and energy levels, an approach that can be generalized to both integrable multi-dimensional systems (EKB torus quantization \cite{Keller}), as well as to systems with multi-component wave functions\cite{zhang,delplace,oz12,kat1,kat2,korm08,wilk84}. In the latter case, however, the multi-component wavefunction matching inherent in the WKB approach rapidly becomes prohibitively cumbersome as the number of components increases.

The Gutzwiller trace formula circumvents this matching problem by giving a direct expression for the semi-classical density of states. It is, furthermore, valid for systems with a non-integrable (i.e., chaotic) classical limit. While generalizations of the trace formula for the multi-component case have been presented for specific matrix Hamiltonians such as the relativistic Dirac Hamiltonian\cite{Keppeler-Bolte,Kepp03,bolt99,Bolte2004,B,Kbook} and the Dirac-Weyl Hamiltonian of graphene\cite{Carmier-ullmo,diet13}, a general multi-component version of the trace formula has not been considered. In the solid state theory context, however, a semi-classical method applicable to \emph{arbitrary matrix Hamiltonians} would be extremely useful. This is so for two reasons. Firstly, in many systems, for instance few layer graphenes, topological insulators, and semiconductors, one encounters multi-band effective Hamiltonians and hence multicomponent wavefunctions. Secondly, and perhaps most importantly, in the emerging class of low dimensional materials one very often encounters structural deformations occurring on length scales far in excess of the lattice constant. Such deformations are very difficult to treat fully quantum mechanically (due to the huge unit cell sizes involved) while at the same time presenting a natural case for a semi-classical treatment (due to the slowly varying spatial deformation). Examples include flexural ripples in graphene\cite{fas07}, rotational stacking faults in few layer graphene systems\cite{lop07,shall10,shall13,shall16}, the recently discovered partial dislocation networks in few layer graphenes\cite{ald13,butz14,kiss15}, and graphene nanostructures for which interesting semi-classical work already exists\cite{Worm11}. A general semi-classical approach for treating such systems thus has the potential of providing a very useful investigative tool.

The purpose of the present paper is therefore twofold: (i) to generalize the Gutzwiller trace formula to the case of arbitrary matrix-valued Hamiltonians and (ii) to demonstrate that the semi-classical approach yields an \emph{accurate and tractable} scheme for the treatment of deformations in graphene based systems. To that end, we will first focus on fundamental theory and some simple applications, and in the final part of the paper consider application of the theory to a realistically complex example of a deformation in bilayer graphene, a one dimensional strain moir\'e.

Let us briefly outline the differences between the matrix-valued case and the scalar case within a semi-classical treatment. At $\mathcal{O}(\hbar^0)$ a scalar Schr\"odinger equation reduces to the Hamilton-Jacobi equation of classical mechanics. For a $n\times n$ Hamiltonian, however, there are two important differences in the $\mathcal{O}(\hbar^0)$ classical structures. Firstly, we obtain $n$ Hamilton-Jacobi equations, some of which may be identical. This situation arises, for instance, in the $\hbar\to0$ limit of the Dirac equation\cite{Keppeler-Bolte}, where each of the two Hamilton-Jacobi equations (for particles and antiparticles) is twice degenerate - in the limit $\hbar\to0$ there is no spin and hence one obtains two pairs of degenerate equations. As for the case of ordinary perturbation theory, such degenerate cases require a special treatment. Secondly, the Hamiltonian phase space is found to be endowed with a classical Berry curvature arising from anholonomy in the transport of the eigenvectors $\hat{V}_\alpha$ of $H(\hat{\bp}\to\bp,\hat{\bq}\to\bq)$ around the classical orbits. In contrast to the scalar case, for multi-band quantum Hamiltonians the semi-classical particle types have ``internal structure'' (pseudospin structure in the case of graphene, for example) and it is this that is encoded in the classical eigenvectors of $H(\bp,\bq)$; the classical Berry curvature is thus a direct consequence of the internal degrees of freedom of the underlying quantum system.

At $\mathcal{O}(\hbar)$ we obtain an equation for the amplitude that is transported along the classical orbits described by the Hamilton-Jacobi equations of motion. In contrast to the scalar case, for a matrix Hamiltonian this amplitude acquires an addition phase resulting from the $\mathcal{O}(\hbar^0)$ Berry curvature which, for non-degenerate systems, consists of (i) a geometric phase and (ii) a dynamical phase acquired by transport of the classical particle through this Berry curvature. The geometric part of this phase depends only on individual Hamilton-Jacobi orbits, while the dynamical phase couples all the $n$ Hamilton-Jacobi systems. For degenerate systems one finds the non-Abelian analogue of the Berry curvature and phase. The existence of a dynamical phase related to the Berry curvature (in addition to the usual geometric phase) is somewhat unusual, and reflects the fact that we have a transport equation for the amplitude function and not the semi-classical vectors $\hat{V}_\alpha$.

Our approach differs from that taken by Carmier and Ullmo\cite{Carmier-ullmo} in their semi-classical study of graphene and 2-band bilayer graphene in that their description begins with the energy dependent Greens function, which then enforces a complex matching procedure for the boundary condition of this function; as in the case of the WKB method, this becomes prohibitively cumbersome for a general matrix valued Hamiltonian. Following Bolte and Keppeler\cite{Keppeler-Bolte} we implement the boundary conditions for the Greens functions at the level of the time dependent Greens functions, which allows an elegant solution that circumvents the boundary matching problem.

The procedure leading from an arbitrary matrix valued Hamiltonian to the density of states is presented as an explicit set of steps, and we apply it to a number of cases where the exact solution is known: the Landau spectra of graphene, a 4-band model of bilayer graphene, and silicene. As one would expect, the agreement between the exact and the semi-classical results becomes considerably degraded at low energies, and in particular for bilayer graphene the zero mode found in the exact solution is not captured within the semi-classical approximation (and is, of course, also not captured in the 2-band approximation to this problem\cite{Carmier-ullmo}). It is therefore by no means obvious that a semi-classical approach is suitable for graphene based systems with slowly spatial deformations, as one is always interested in the low energy behaviour.

To explore this situation more closely we consider a realistic example of such a deformation: a one dimensional strain moir\'e in bilayer graphene, which serves as an instructive analogue of the graphene twist bilayer\cite{san12}, two mutually rotated layers of graphene. The twist bilayer displays extraordinarily rich electronic structure in small angle limit \cite{lop07,shall10,bist11,shall13,kaz14,shall16} (i.e., as the moir\'e length becomes large), and in particular a strong electron localization on the AA stacked regions of the lattice. We present an analytical semi-classical analysis of the strain moir\'e, finding that: (i) at the Dirac point the action is orders of magnitude larger than $\hbar$ when the moir\'e length becomes large compared to the lattice constant - thus validating the semi-classical approach - and (ii) that the electron localization is driven by the existence of a \emph{semi-classical potential well} centered at the AA spots, that arises from the stacking potential in the quantum Hamiltonian. We should stress that this ``potential well'' picture, which provides a very natural description of charge localization, is fundamentally semi-classical: no such ``potential well'' could localize quantum mechanically due to the Klien paradox which prevents localization of electrons in graphene by a scalar potential. We furthermore find an analytical form for the semi-classical density of states arising from electrons trapped in this potential well, which we show to be in almost perfect agreement with exact quantum mechanical calculations. This demonstrates both that the semi-classical approach provides a valid tool for investigating slow deformations in few layer systems, as well as the insight it can bring into the physics of these rather complex systems. It should be stressed that treating such a system on the basis of either the standard WKB approach (in principle possible as we have an effective one-dimensional system), or that of Ref.~\onlinecite{Carmier-ullmo}, could not be contemplated due to the extraordinarily complexity of the matching procedure that would be involved.


\section{Semi-classics for matrix valued Hamiltonians}

We consider Schr\"odinger's equation for the Greens function $G(\br,\br^\prime,t)$ with a \emph{matrix-valued} Hamiltonian $\hat H$

\begin{equation}
 i\hbar \partial_t G(\br,\br^\prime,t)=\hat H(-i\hbar\partial_i,x_i)G(\br,\br^\prime,t)
\label{sg1}
\end{equation}
and the boundary condition

\begin{equation}
 G(\br,\br^\prime,0)=\mathbb{1}_n\delta(\br-\br^\prime).
\label{boundary}
\end{equation}
We assume that $\hat H$ is analytic in $-i\hbar\partial_i$, or more precisely that for all smooth test functions $\Phi$ the matrix $\hat H(\partial_i\Phi,x_i)$ is analytical in $\partial_i\Phi$. We further assume that the Hamiltonian may be decomposed as $\hat H=\hat H_D(-i\hbar\partial_i )+\hat H_x(x_i)$, in which $\hat H_D$ is a matrix containing only derivatives (such as momentum operators) and $\hat H_x$ contains only coordinate functions such as potentials. This restriction is easily lifted without qualitatively changing what follows, however it covers most physical situations and simplifies the analysis so we retain it. Finally, we introduce the following convention: the \emph{primed} coordinates are \emph{initial} coordinates with the \emph{non-primed} ones the \emph{final} coordinates on a classical trajectory. \\

Let us first derive an expression for the time dependent Greens function using a generalization of the ansatz provided by Bolte and Keppeler \cite{Keppeler-Bolte}. If $\hat H$ is a $n\times n$ matrix we search for the \emph{time-dependent Greens function} of the form

\begin{eqnarray}
 G(\br,\br^\prime,t) & = & \frac{1}{(2\pi\hbar)^d}\int d^dp^\prime \sum_\alpha
 \hat V_\alpha(t,\br,\bp^\prime) \label{BKan}
 \\
 & \times & \hat f_\alpha(t,\br,\br^\prime,\bp^\prime)\hat V_\alpha^\dagger(0,\br^\prime,\bp^\prime) e^{\frac{i}{\hbar}\Phi_\alpha(\br,\br^\prime,\bp^\prime,t)} \nonumber.
\end{eqnarray}
This ansatz is completely general and could describe the exact solution to Eq.~(\ref{sg1}), however the $e^{\frac{i}{\hbar}\Phi_\alpha(\br,\br^\prime,\bp^\prime,t)}$ term allows a WKB-like expansion in orders of $\hbar$ and, due to the integral in Eq.~(\ref{BKan}), one can implement the boundary condition Eq.~\eqref{boundary} by requiring

\begin{equation}
 \begin{split}
 \hat f_\alpha(t=0)=1\\
\Phi_\alpha(t=0)=\bp^\prime (\br-\br^\prime)\\
\sum_\alpha \hat V_\alpha(0,\br^\prime,\bp^\prime) V_\alpha^\dagger(0,\br^\prime,\bp^\prime)=\mathbb{1}_n
 \end{split}.
\label{initialcond0}
\end{equation}
We now assume that $\hat V_\alpha=\hat V_\alpha^{(0)}+\hbar \hat V_\alpha^{(1)}+...$ is analytical in $\hbar$ and insert Eq.~\eqref{BKan} into the Schr\"odinger equation Eq.~\eqref{sg1} and collect terms of $\mathcal{O}(\hbar^0)$. This procedure leads to the \emph{zero eigenvalue condition}

\begin{equation}
 ( H(\partial_\mu \Phi_\alpha,x_i)+\partial_t\Phi_\alpha)\hat V_\alpha^{(0)}=0.
\label{sgapprox1}
\end{equation}
(We may drop the sum and integral in Eq.~\eqref{BKan} as the equation is linear.) There are evidently $n$ solutions to this equation which occur when the $\hat V_\alpha^{(0)}$ are equal to the $n$ orthonormal eigenvectors of $H(\partial_\mu \Phi_\alpha,x_i)$. Denoting the corresponding eigenvalue by $H_\alpha$, we find that the zero eigenvalue equation implies the existence of $n$ Hamilton-Jacobi equations

\begin{equation}
H_\alpha=-\partial_t \Phi_\alpha.
\label{HJeq}
\end{equation}
from which we then immediately identify for system $\alpha$ the following objects: $H_\alpha$ is a classical scalar Hamiltonian, the real scalar function $\Phi_\alpha(\br,\br^\prime,\bp^\prime,t)$ a classical action, and $p_\mu^\alpha = \partial_\mu \Phi_\alpha$ the corresponding canonical momentum. Thus a general $n\times n$ quantum Hamiltonian leads to $n$ separate classical systems at order $\mathcal{O}(\hbar^0)$. The full classical information is contained in the $n$ Hamilton-Jacobi equations (\ref{HJeq}) and the $n$ vectors $\hat V_\alpha^{(0)}$.


The situation becomes more complicated if the matrix $H(p_i,x_i)$ has degenerate eigenvalues. In this circumstance recourse to degenerate perturbation theory may be circumvented following the scheme of Bolte and Keppeler, in which the $m_\alpha$ orthogonalized eigenvectors corresponding to the degenerate eigenfunction $H_\alpha$ are combined into an $m_\alpha \times n$ ``full degenerate eigenvector'' that evidently satisfies $\hat V_\alpha^{(0)\dagger}H(p_i,x_i)\hat V_\alpha^{(0)}=H_\alpha\mathbb{1}_{m_\alpha}$. With the initial conditions

\begin{equation}
 \begin{split}
 \hat f_\alpha(t=0)=\mathbb{1}_m\\
\Phi_\alpha(t=0)=\bp^\prime (\br-\br^\prime)\\
\sum_\alpha \hat V_\alpha(0,\br^\prime,\bp^\prime) \hat V_\alpha^\dagger(0,\br^\prime,\bp^\prime)=\mathbb{1}_n
 \end{split},
\label{initialcond2}
\end{equation}
the boundary condition of the Green's function equation is satisfied and the calculations thereafter proceed in a formally identical way for both the degenerate and non-degenerate cases, with the difference of course that the rank of the various objects in the theory differs between the two cases. \\


We now consider the $\mathcal{O}(\hbar)$ corrections to the $n$ Hamilton-Jacobi equations; sewing ``quantum flesh on classical bones'' to use the evocative phrase of Berry and Mount \cite{ber72}. The derivative operators $-i\hbar\partial_i$ in $\hat H_D$ are substituted by 
the canonical momenta $p_\mu^\alpha$ upon acting on $e^{\frac{i}{\hbar}\Phi_\alpha}$. To generate the $\mathcal{O}(\hbar)$ terms from $\hat H_D$ we should allow every matrix element $\left[\hat H_D\right]_{ij}$ to act on the exponential until only first order operator terms remain. For example, the term $i\hbar^3\partial_i\partial_j\partial_k$ would generate $-i\hbar p_{i \alpha}p_{j,\alpha}\partial_k+(k\leftrightarrow j)+(k\leftrightarrow i)$. The resulting matrix operator $-i\hbar H^1_D$ is therefore linear in first order derivatives. Considering $-i\hbar H^1_D$ as a perturbation, we obtain in first order perturbation theory the $\mathcal{O}(\hbar)$ equation

\begin{equation}
\hat V_{\alpha}^{(0)\dagger}\left[ H_D^1+\partial_t\right]\hat V_\alpha^{(0)}\hat f_\alpha(\br)=0,
\label{Teq1}
\end{equation}
which, as we shall now show, is a \emph{transport equation} for $f_\alpha(\br)$ along the classical trajectories governed by the $H_\alpha$ Hamiltonian. (Note that substitution of Eq.~\eqref{BKan} into the Schr\"odinger equation Eq.~\eqref{sg1}, collecting terms of $\mathcal{O}(\hbar)$, and pre-multiplying by $V_{\alpha}^{(0)\dagger}$ also leads to Eq.~\ref{Teq1}.) In the case of degenerate Hamilton-Jacobi systems $f_\alpha(\br)$ will be an $m_\alpha \times m_\alpha$ matrix; in the non-degenerate case $f_\alpha(\br)$ is a scalar function. It is not apparent from Eq.~(\ref{Teq1}) that the $n$ formally independent classical systems found at $\mathcal{O}(\hbar)$ are coupled at $\mathcal{O}(\hbar)$ although, as we shall see, different particle types are indeed coupled by the transport equation.

To simplify the notation we now define

\begin{equation}
\begin{split}
 H_{cl}(p_i,x_i):=H(p_i,x_i)+\partial_t \Phi_\alpha\\
\lambda_\alpha=\partial_t\Phi_\alpha+H_\alpha
\end{split}
\label{Hcl}
\end{equation}
such that Eq.~\eqref{sgapprox1} and Eq.~\eqref{HJeq} take the form $H_{cl}\hat V_\alpha=0$ and $\lambda_\alpha=0$ respectively. 

We now demonstrate that Eq.~\eqref{Teq1} may be written in the form of a transport equation familiar from the scalar case. We first note a rather obvious relation connecting $H^1_D$ and $H_{\mathrm{cl}}$ namely:

\begin{equation}
 \left[H_D^1\right]_{lm}=\sum_{k=1}^d\partial_{p_k}\left[H_{cl}\right]_{lm}\partial_k.
\label{hDd}
\end{equation}
This identity follows from deployment of the Leibnitz rule on the right hand side that generates all terms from $H(p_i,x_i)$ with one momentum variable removed and replaced by the operator $\partial_k$, exactly the definition of $H_D^1$ on the left hand side. The identity Eq.~\eqref{HDFin} generalizes a similar relation used by Carmier and Ullmo\cite{Carmier-ullmo} in their semi-classical treatment of single layer graphene and the two band model of bilayer graphene. In accordance with convention we set $\partial_t=\partial_0$,  $\partial_E=\partial_{p_0}$ and make use of the sum convention letting the sum run from  $\mu=0,...,d$, where $d$ is the space dimension. We may then compactly write

\begin{equation}
 H_D^1+\partial_t=\partial_{p_\mu}H_{cl}\partial_\mu.
\label{HDFin}
\end{equation}
and using this result can write Eq.~(\ref{Teq1}) as

\begin{equation}
 \left[\hat V_\alpha^{(0)\dagger}\left[\partial_{p_\mu}H_{cl}\right]\partial_\mu\hat V_\alpha^{(0)}
+\hat V_\alpha^{(0)\dagger}\left[\partial_{p_\mu}H_{cl}\right]\hat V_\alpha^{(0)}\partial_\mu\right]\hat f_\alpha(\br)=0.
\label{eq212}
\end{equation}
Since we do not intend to calculate the higher order corrections to $\hat V_\alpha^{(0)}$ from now on we shall drop the index $(0)$ at all vectors $V_\alpha$.

Now, following Carmier and Ullmo \cite{Carmier-ullmo} we split $\hat V_\alpha^\dagger\partial_{p_\mu}H_{cl}\partial_\mu\hat V_\alpha$ into
a hermitian and an anti-hermitian part. We denote the \emph{antihermitian} part $iM_\alpha$:

\begin{equation}
iM_\alpha=\mathrm{Antiherm}\left(\hat V_\alpha^\dagger\partial_{p_\mu}H_{cl}\partial_\mu\hat V_\alpha\right)
\label{Malpha111}
\end{equation}
(Note that in the case of an $m_\alpha$-fold degeneracy $M_\alpha$ is a $m_\alpha\times m_\alpha$ matrix.) This term includes a semi-classical analogue of a Berry phase, as explained in detail later.

By repeatedly applying the Leibnitz rule and using the fact that $\partial_{p_\mu} H_{cl}$ is Hermitian and $\partial_\mu\partial_{p_\mu}H_{cl}=0$ (recall our assumption of no coordinate functions in front of derivatives in the Hamiltonian) we see that the Hermitian part of $\hat V_\alpha^\dagger\partial_{p_\mu}H_{cl}\partial_\mu\hat V_\alpha$ takes the following simple form
\begin{equation}
\mathrm{Herm}({\hat V_\alpha}^\dagger [\partial_{p_\mu}H_{cl}]\partial_\mu \hat V_\alpha) = 
 \frac{1}{2}\partial_\mu(\partial_{p^\alpha_\mu}\lambda_\alpha).
\label{transpeqrealpart}
\end{equation}
Where we have also used the Hellmann-Feynman theorem $\hat V_\alpha^\dagger\partial_\gamma H_{cl}V_\alpha=\partial_\gamma \lambda_\alpha$. Applying the Hellmann-Feynman theorem to the second part of Eq.~\eqref{eq212} we find that the equation simplifies to

\begin{equation}
 \left[\frac{1}{2}\partial_\mu \left( \partial_{p^\alpha_\mu}\lambda_\alpha\right)+
\left( \partial_{p^\alpha_\mu}\lambda_\alpha\right)\partial_\mu+iM_\alpha\right]\hat f_\alpha(\br)=0.
\label{eq2124}
\end{equation}
From Hamilton's equations we have $\left( \partial_{p^\alpha_\mu}\lambda_\alpha\right)=\dot x_\mu$ with $\dot x_0=1$ and so Eq.~\eqref{eq2124} reduces to the \emph{transport equation}

\begin{equation}
 \left[\frac{1}{2}\partial_\mu \left( \partial_{p^\alpha_\mu}\lambda_\alpha\right)+\frac{d}{dt}+iM_\alpha\right]\hat f_\alpha(\br)=0,
\label{eq2123}
\end{equation}
which is similar to the scalar case but \emph{with an additional matrix $M_\alpha$}. It should be stressed that $M_\alpha$ appears only for matrix-valued Hamiltonians. All terms in Eq.~\eqref{eq2123} are expressed in terms of the classical orbits obtained from the order $\mathcal{O}(\hbar^0)$ approximation.

Equation~\eqref{eq2123} must now be integrated along the classical orbits such that $f_\alpha(\br)$ is ``transported'' along these orbits. The first two terms of this expression are scalars, and hence the ansatz $\hat f_\alpha=g_\alpha(\br)\hat h_\alpha(\br)$, where $g_\alpha$ 
is a scalar function that solves an auxiliary equation 
$\left[\frac{1}{2}\partial_\mu \left( \partial_{p^\alpha_\mu}\lambda_\alpha\right)+\frac{d}{dt}\right]g_\alpha(\br)=0$, is sensible. The term $\hat h_\alpha(\br)$ is an $m_\alpha\times m_\alpha$ matrix in case of an $m_\alpha$-fold degeneracy for $H_\alpha$. The expression for $g_\alpha(\br)$ with initial condition $\left.g_\alpha(\br)\right|_{t=0}=1$ is known from the scalar case (see, for example, Refs.~\onlinecite{Maslov1,Maslov2}):

\begin{equation}
 g_\alpha(\br)=\sqrt{\mathrm{det}\left(\frac{\partial(\bp_\alpha,t)}{\partial(\bp^\prime,t)}\right)}.
\label{galpha}
\end{equation}
Inserting $\hat f_\alpha=g_\alpha(\br)\hat h_\alpha(\br)$ into Eq.~\eqref{eq2123} yields 

\begin{equation}
 \left[\frac{d}{dt}+iM_\alpha\right]\hat h_\alpha(\br)=0,
\label{insertion}
\end{equation}
which is then solved by a \emph{time-ordered exponential} (matrices $M_\alpha$ at different times will in general not commute).


We now have all ingredients required to calculate the time-dependent Greens function Eq.~\eqref{BKan}. The remaining steps of the derivation now follow closely the scalar case, and we will, therefore, be brief in the presentation. From the initial condition Eq.~\eqref{initialcond0} for the classical actions $\Phi_\alpha$ it follows that 
$\Phi_\alpha=\phi_\alpha(\br,\bp^\prime,t)-\bp^\prime\br^\prime$\cite{Keppeler-Bolte}, where $\phi_\alpha$ are as yet unknown functions. We evaluate Eq.~\eqref{BKan} with the stationary phase approximation to find

\begin{eqnarray}
 G(\br,\br^\prime,t) & \approx & \frac{1}{(2\pi i\hbar)^{\frac{d}{2}}}\sum_{\alpha,\gamma_t} 
 g_{\alpha,\gamma_t}
 \hat V_{\alpha,\gamma_t}(\br,t) \\
 & \times & \hat h_{\alpha,\gamma_t}(\br,\br^\prime,t)
 \hat V_{\alpha,\gamma_t}^\dagger(\br^\prime,0)
 e^{\frac{i}{\hbar} S_{\alpha,\gamma_t}(\br,\br^\prime,t)-i\frac{\pi}{2}\nu_{\alpha,\gamma_t}} \nonumber
\end{eqnarray}
with

\begin{equation}
g_{\alpha,\gamma_t} = \sqrt{\left|-\mathrm{det}\left(\frac{\partial(\bp_{\alpha,\gamma_t},t)}{\partial(\br^\prime,t)}\right)\right|}.
\label{timedep-GF}
\end{equation}
Here the stationary phase applied to the $\bp^\prime$ integral has enforced the sum over $\gamma_t$ to be over all classical paths leading from $\br^\prime$ to $\br$ within the time interval $t$. The actions $S_{\alpha,\gamma_t}(\br,\br^\prime,t)$ are the classical actions that solve the Hamilton-Jacobi equations with this requirement. Additionally, $\nu_{\alpha,\gamma_t}$ is the path specific time-dependent Maslov index (or Morse index) that arises from taking the absolute value in the expression for $g_{\alpha,\gamma_t}$. The Morse index accounts for the sign changes of the determinant under the square root, which occurs when the Lagrangian manifold corresponding to the dynamics of the Hamiltonian $H_\alpha$ develops a fold (see, for example, the book of Cvitanovic\cite{cvitanovic}). In a one dimensional problem this simply corresponds to the turning points of the classical path at which the Greens function picks up a phase $\frac{\pi}{2}$.

We now determine the retarded energy-dependent Greens function

\begin{equation}
 G(\br,\br^\prime,E)=\lim_{\epsilon\to 0}\frac{1}{i\hbar}\int_0^\infty dt\,G(\br,\br^\prime,t)e^{\frac{i}{\hbar}(E-i\epsilon)t}.
\label{endepgf}
\end{equation}
within the stationary phase approximation. The calculation is again almost identical to the scalar case and we thus, for brevity of presentation, refer the reader to the standard literature for the scalar Gutzwiller formula (see, for example, Ref.~\onlinecite{cvitanovic}) and quote the result of the stationary phase approximation for the matrix valued case:

\begin{eqnarray}
 G(\br,\br^\prime,E) & \approx & \frac{1}{i\hbar(2\pi i\hbar)^{(d-1)/2}}\sum_{\alpha,\gamma_E}g_{\alpha,\gamma_E} \label{endepgf0} \\
 &\times&\hat V_{\alpha,\gamma_E}(\br,\br^\prime) \hat h_{\alpha,\gamma_E}(\br,\br^\prime) \hat V_{\alpha,\gamma_E}^\dagger(\br^\prime,\br^\prime) \nonumber \\
 &\times& e^{\frac{i}{\hbar} S_{\alpha,\gamma_E}(\br,\br^\prime,E)-i\frac{\pi}{2}\nu_{\alpha,\gamma_E}}\nonumber,
\end{eqnarray}
where

\begin{equation}
 g_{\alpha,\gamma_E}=\sqrt{\left|-\frac{1}{\dot r^\parallel_{\alpha,\gamma_E},\dot r^{\parallel\prime}_{\alpha,\gamma_E}}
\mathrm{det}\left(\frac{\partial^2 S(\br,\br^\prime,E)}{\partial\br^\perp\partial\br^{\perp\prime}}\right)\right|},
\label{endepgf2}
\end{equation}
and where the stationary phase condition now sets the summation to be on the \emph{classical energy shell}. In this expression $S(\br,\br^\prime,E)$ is now the energy-dependent action, and $\gamma_E$ are all classical paths connecting $\br$ and $\br^\prime$ at energy $E$. As is customary within a semi-classical formulation we have chosen a coordinate system with one axis parallel ($\parallel$) to the trajectory and the other coordinates perpendicular ($\perp$) to the trajectory. Similar to Eq.~\eqref{endepgf}, $\nu_{\alpha,\gamma_E}$ is the \emph{Maslov index} and counts the sign changes of the expression under the absolute value in $g_{\alpha,\gamma_E}$. 

We can now calculate the density of states simply by taking a trace over positions and over the matrix structure

\begin{equation}
 d(E)=-\frac{1}{\pi}\lim_{\epsilon\rightarrow+0}\mathrm{Im}\left\{\mathrm{tr}\left[G(\br,\br^\prime,E+i\epsilon)\right]\right\}.
\label{densityofstates}
\end{equation}
The calculation of the position trace is, once again, identical to the scalar case which, following the standard procedure, evaluated in a stationary phase approximation for perpendicular coordinates and without any approximation (but assuming isolated orbits) for the parallel coordinates. This yields for the density of states

\begin{eqnarray}
d_{\mathrm{osc}}(E) & = & \sum_{\alpha,\gamma^\circ_E}\mathrm{Im}\Bigg(\frac{iT^p_{\alpha,\gamma^\circ_E}}{\hbar\pi}
\mathrm{tr}\left(\hat h_{\alpha,\gamma_E^\circ}\right) \nonumber \\
&\times&\frac{e^{\frac{i}{\hbar}S_{\alpha,\gamma^\circ_E}
-i\frac{\pi}{2}\nu_{\alpha,\gamma_E^\circ}}}{\sqrt{\left|\mathrm{det}\left(\mathbb{J}_{\alpha,\gamma^\circ_E}
-\mathbb{1}_{2(d-1)}\right)\right|}} \Bigg)
\label{denstates}
\end{eqnarray}
where the position trace restricts the summation to the closed orbits and the stationary phase condition picks up only periodic orbits $\gamma_E^\circ$, and where the monodromy matrix $\mathbb{J}$ is given as

\begin{equation}
\mathbb{J}_{\alpha,\gamma^\circ_E}=\left.\frac{\partial(p_{\alpha,\gamma^\circ_E}^{\perp},x_{\alpha,\gamma^\circ_E}^{\perp})}
{\partial(p_{\alpha,\gamma^\circ_E}^{\perp\prime},x^{\perp\prime})}\right|_{(x_{\alpha,\gamma^\circ_E}^{\perp},p_{\alpha,\gamma^\circ_E}^{\perp})
 \in\gamma^\circ_E},
 \label{monodrom}
\end{equation}
and which describes the stability of an orbit with respect to small deviations of the initial positions and momenta. It is evaluated for the orbit $\gamma_E^\circ$ and is independent of the point on the orbit which we choose for its evaluation\cite{cvitanovic}. The \emph{Maslov index} is $\nu_{\alpha,\gamma_E^\circ}$ and it is important to note that, in general, it is not the same as the Maslov index in Eq.~\eqref{endepgf2}. For one dimensional systems, however, these indices coincide. Lastly $T^p_{\alpha,\gamma^\circ_E}$ is the \emph{time} needed to traverse the \emph{primitive orbit} (primitive means traversing the orbit only once). Note that this result is valid only for the extended orbit contributions. For the short orbits the exponential is not a fast oscillating function and the stationary phase approximation is inadmissible. Equation~\eqref{denstates} provides the so-called oscillatory part of the density of states.

For the non-degenerate case the $M$ phase is a scalar, and thus Eq.~(\ref{insertion}) may be immediately solved to yield $h_{\alpha \gamma_E^\circ}=e^{i\int_0^{T_{\alpha,\gamma^\circ_E}} dt\,M_{\alpha}}$, leading to a simpler Gutzwiller formula

\begin{eqnarray}
d_{\mathrm{osc}}(E)&=&\sum_{\alpha,\gamma^\circ_E}\Bigg[\frac{T^p_{\alpha,\gamma^\circ_E}}{\hbar\pi} \label{simplf}
\\
&\times & \frac{\cos\left(\frac{1}{\hbar}
S_{\alpha,\gamma^\circ_E}-\frac{\pi}{2}\nu_{\alpha,\gamma_E^\circ}+\int_0^{T_{\alpha,\gamma_E^\circ}}dtM_\alpha\right)}
{\sqrt{\left|\mathrm{det}\left(\mathbb{J}_{\alpha,\gamma^\circ_E}-\mathbb{1}_{2(d-1)}\right)\right|}}\Bigg],\nonumber
\end{eqnarray}
an expression that, in fact, differs from the scalar formula only by the addition of the $M_\alpha$ phase.


\subsection{Relation between the semi-classical phase and the Berry phase}

The semi-classical $M_\alpha$ phase appearing in the matrix generalization of the Gutzwiller equation has been discussed by Carmier and Ullmo\cite{Carmier-ullmo} in the context of graphene and a 2-band model of bilayer graphene. These authors concluded that for quantum Hamiltonians with no mass term, but arbitrary field $V(\br)$, this phase was exactly the classical analogue of the adiabatic Berry phase. Here we wish to show that this semi-classical phase, that would seem to appear only at $\mathcal{O}(\hbar)$, is in fact a manifestation of a deeper $\mathcal{O}(\hbar^0)$ classical structure in the semi-classical theory of matrix Hamiltonians. 
The full form of this phase for the $\alpha$ HJ system is given by

\begin{equation}
 \int_0^T \!\!\!dt\,\, M_\alpha = \int_0^T \!\!\!dt\,\, \Im V^\dagger_\alpha \left[\partial_{p_\mu} H_{cl}\right]\partial_\mu V_\alpha
 \label{SCPH}
\end{equation}
where for simplicity of notation we consider the case of non-degenerate Hamilton-Jacobi equations (the generalization is straightforward). Switching from 4-vector notation to separate space and time derivatives we have for $M_\alpha$

\begin{equation}
  V^\dagger_\alpha \left[\partial_{p_\mu} H_{cl}\right]\partial_\mu V_\alpha = V^\dagger_\alpha \left[\partial_{p_i} H\right]\partial_i V_\alpha
 + V^\dagger_\alpha \partial_t V_\alpha
\end{equation}
If we insert into the right hand side of this equation $H = \sum_\beta H_\beta V_\beta V_\beta^\dagger$ we then find for the integrand

\begin{equation}
 \Im\left\{ \dot{x}_\mu V_\alpha^\dagger \partial_i V_\alpha + V^\dagger_\alpha \partial_t V_\alpha
 + \sum_\beta H_\beta V_\alpha^\dagger \partial_{p_i}(V_\beta V_\beta^\dagger) \partial_i V_\alpha \right\},
 \label{YO}
\end{equation}
the first two terms of which are evidently the total time derivative

\begin{equation}
 \Im V_\alpha^\dagger d_t V_\alpha,
\end{equation}
i.e., represent a Berry phase that depends only on the geometry of the classical path (as the total time derivative allows us to eliminate time from the integral $\int_0^T dt V_\alpha^\dagger d_t V_\alpha$ by a change of variables). We write this, for a general $2d$ parameter space ($d$ is the dimension of space) and using the notation $\bR = (x_1,x_2,\hdots,x_d,p_1,p_2,\hdots,p_d)$, as

\begin{equation}
 \int_\Sigma\frac{1}{2} F_{\alpha\mu\nu} dR^\mu\wedge dR^\nu
\end{equation}
where $F_{\alpha\mu\nu} = \partial_\mu A_{\alpha\nu} - \partial_\nu A_{\alpha\mu}$ is the Berry curvature tensor,  $A_{\alpha\mu} = i V_\alpha^\dagger \partial_\mu V_\alpha$ the Berry connection for the $\alpha$ Hamilton-Jacobi system, and $\Sigma$ a hyper-surface in the $2d$ Hamiltonian phase space. Note that the Greek indices $\mu$, $\nu$ now run over the $2d$ dimensions of phase space in the vector $\bR$.

What is the third term of Eq.~\eqref{YO}? Evaluation of the derivative $\partial_{p_i}(V_\beta V_\beta^\dagger)$ and insertion of the identity operator yields

\begin{equation}
 \Im \sum_{\beta} (H_\beta-H_\alpha) (V_\alpha^\dagger \partial_{p_i} V_\beta) (V_\beta^\dagger \partial_i V_\alpha)
 \label{DYPH}
\end{equation}
(note that the $\alpha=\beta$ term is identically zero in this sum).
Using $\Im X = -i(X-X^\ast)/2$ we may write this as

\begin{eqnarray}
&& -\frac{i}{2}\sum_{\beta} (H_\beta-H_\alpha) \Big[
 (V_\alpha^\dagger \partial_{p_i} V_\beta) (V_\beta^\dagger \partial_i V_\alpha)\nonumber \\
 &-&
 (V_\alpha^\dagger \partial_{i} V_\beta) (V_\beta^\dagger \partial_{p_i} V_\alpha)
 \Big]
 \label{EQPH1}
\end{eqnarray}
which has some resemblance to the antisymmetric structure of the Berry curvature tensor,

\begin{eqnarray}
 F_{\alpha\mu\nu} & = & i\sum_{\beta}\Big( (V_\alpha^\dagger \partial_\nu V_\beta) (V_\beta^\dagger \partial_\mu V_\alpha) \nonumber \\
 &-&
 (V_\alpha^\dagger \partial_\mu V_\beta) (V_\beta^\dagger \partial_{\nu} V_\alpha)\Big).
\end{eqnarray}
Eq.~(\ref{EQPH1}), however, contains only terms diagonal in dimension and that have mixed $p$ and $x$ derivatives, while the curvature tensor contains all possible combinations of these indices. However, by contracting the curvature tensor with the symplectic matrix $\Omega$ we find

\begin{eqnarray}
 \frac{1}{4} \Omega^{\mu\nu} F_{\alpha\mu\nu}& =& \frac{i}{2} \sum_{\beta} \Big[(V_\alpha^\dagger \partial_{p_i} V_\beta) (V_\beta^\dagger \partial_i V_\alpha) \nonumber \\
 &-&
 (V_\alpha^\dagger \partial_{i} V_\beta) (V_\beta^\dagger \partial_{p_i} V_\alpha)
 \Big]
\end{eqnarray}
which is, apart from the weight factor $(H_\beta-H_\alpha)$, identical to Eq.~(\ref{EQPH1}).

We may therefore express the semi-classical phase as

\begin{eqnarray}
 \int \!\! dt \,M_\alpha &=& \int_\Sigma\frac{1}{2} F_{\alpha\mu\nu} dR^\mu\wedge dR^\nu  \label{SCP} \\
 &-& \frac{1}{4} \int\! dt\,\sum_{\beta} (H_\beta(\bR(t))-E)\Omega^{\mu\nu} F_{\alpha\beta\mu\nu}(\bR(t)) \nonumber
\end{eqnarray}
where we have introduced the $\beta$ state part of the curvature tensor:

\begin{equation}
 F_{\alpha\beta\mu\nu} = \left[(V_\alpha^\dagger \partial_\mu V_\beta) (V_\beta^\dagger \partial_\nu V_\alpha) -
 (V_\alpha^\dagger \partial_\nu V_\beta) (V_\beta^\dagger \partial_\mu V_\alpha)
 \right]
\end{equation}
and used the fact that on the $\alpha$ orbit we have $H_\alpha = E$.

Eq.~\eqref{SCP} has a simple visual interpretation. The first term is the geometric Berry phase while the second term is clearly dynamical, and represents the time integral of the classical particle moving through a Hamiltonian phase space endowed with a Berry curvature. Contraction of the Berry curvature with the symplectic matrix yields a scalar, and in the line integral this is, for each semi-classical vector $\beta$, weighted by the energy separation of the $\beta$ manifold and the $\alpha$ particle orbit: $(H_\beta-E)$. This term evidently encodes the coupling between the $n$ non-degenerate Hamilton-Jacobi systems.

One might wonder how important is the dynamical part of the semi-classical phase. As we will subsequently show, in the case of silicene (which has an explicit mass term) this term in the semi-classical phase considerably improves the accuracy of the theory.

\emph{Structure of the dynamical semi-classical phase}: The structure of the Berry curvature depends, as usual, on the degeneracy structure, in this case of the semi-classical energy manifolds. It is worthwhile exploring this point, and to that end we insert the formula for the matrix element of the derivative of an eigenvector

\begin{equation}
 V_\beta^\dagger(\partial\mu V_\alpha) = \frac{V_\beta^\dagger (\partial_\mu H) V_\alpha}{H_\alpha-H_\beta}
 \label{DER}
\end{equation}
into $F_{\alpha\beta\mu\nu}$ to find

\begin{eqnarray}
  F_{\alpha\beta\mu\nu} & = & -\frac{V_\alpha^\dagger (\partial_\mu H) V_\beta V_\beta^\dagger (\partial_\nu H) V_\alpha}{(H_\beta-H_\alpha)^2}\nonumber\\
  &+&
 \frac{V_\alpha^\dagger (\partial_\nu H) V_\beta V_\beta^\dagger (\partial_\mu H) V_\alpha}{(H_\beta-H_\alpha)^2}
\end{eqnarray}
The structure of $F_{\alpha\beta\mu\nu}$ is thus dominated by degeneracies amongst the classical eigenvalues. The geometric part of the semi-classical phase, of course, depends on the global structure of the Berry curvature $F_{\alpha\mu\nu} = \sum_\beta F_{\alpha\beta\mu\nu}$. This structure is, however, equally important for understanding the dynamical phase. In particular, if the Berry curvature is a $\delta$-function - the case for which the Berry phase is ``topological'', i.e., depends only on the winding number around the pole - then any classical trajectory that does not pass through such a source will have zero for the line integral in Eq.~(\ref{SCP}). In other words, \emph{if the semi-classical Berry phase is topological then the semi-classical dynamical phase is zero}. A special case of this is the graphene Dirac-Weyl Hamiltonian in the presence of arbitrary $V(\br)$ but with no mass term $\sigma_z \phi(\br)$. In such a situation the degeneracy at the Dirac point is preserved, the curvature retains the Delta function structure, and hence the semi-classical phase will coincide with the Berry phase, as stated by Carmier and Ullmo\cite{Carmier-ullmo}.


\subsection{Treating the case of orbit degeneracies}

If the Hamiltonian has one or more cyclic coordinates, then the stationary phase approximation for these coordinates cannot be applied. Furthermore, if orbits are not isolated the trace over starting positions is not given by the integral $\oint \frac{dx_\parallel}{\dot x_\parallel}=T^p$ as this clearly assumes a single closed orbit (here $T^p$, as before, denotes the time period of a primitive orbit i.e. the time to travel once around a closed orbit). We consider here the case where only one coordinate $x_n$ is non-cyclic and indicate how this result may be (straightforwardly) generalized. The method presented here differs from that of Carmier and Ullmo\cite{Carmier-ullmo} in that we derive a solution constructively, beginning at the level of the transport equation, whereas in Ref.~[\onlinecite{Carmier-ullmo}] the treatment of orbit degeneracies is performed post a general solution of the transport equation. The two methods are, however, equivalent.

For the case of a single non-cyclic coordinate the transport equation takes on the much simpler form 

\begin{equation}
 \left(\frac{1}{2}\partial_n\partial_{p_n}+\frac{d}{dt}+iM_\alpha\right)\hat f_\alpha=0.
\label{simplertransport}
\end{equation}
with the solution

\begin{equation}
 \hat f_{\alpha}=\sqrt{\frac{\left(\frac{\partial p_n^{\alpha}}{\partial x_n^\prime}\right)}
{    \left( \frac{\partial p_n^{\alpha\prime}}{\partial x_n^\prime}\right)  }}\hat h_\alpha
\label{solution}
\end{equation}
this result for $f_{\alpha}$ gives a new expression for the time-dependent Greens function

\begin{eqnarray}
 G(\br,\br^\prime,t) &\approx& \int\frac{ d p_1...dp_{n-1}}{i^{1/2}(2\pi \hbar)^{d-1/2}}
\sum_{\alpha,\gamma_t} g_{\alpha,\gamma_t}
\hat V_{\alpha,\gamma_t}(\br,\br^\prime) \label{timedep-GFdeg}
\\
&\times&\hat h_{\alpha,\gamma_t}
(\br,\br^\prime)\hat V_{\alpha,\gamma_t}^\dagger(\br,\br^\prime)
e^{\frac{i}{\hbar} S_{\alpha,\gamma_t}
(\br,\br^\prime,t)-i\frac{\pi}{2}\nu_{\alpha,\gamma_t}} \nonumber
\end{eqnarray}
where

\begin{equation}
g_{\alpha,\gamma_t}= \sqrt{\left|-\frac{\partial(p_{\alpha,\gamma_t},t)}{\partial(x_n^\prime,t)}\right|}
\label{timedep-GFdegB}
\end{equation}
while similarly for the energy dependent Greens function we find

\begin{eqnarray}
 G(\br,\br^\prime,E) & \approx &\int\frac{dp_1...dp_{n-1}}{i\hbar(2\pi \hbar)^{d-1}}\sum_{\alpha,\gamma_E}
g_{\alpha,\gamma_E} \label{endepgf2deg} \\
& \times& \hat V_{\alpha,\gamma_E}(\br,\br^\prime)\hat h_{\alpha,\gamma_E}(\br,\br^\prime)
\hat V_{\alpha,\gamma_E}^\dagger(\br^\prime,\br^\prime) \nonumber \\
& \times& e^{\frac{i}{\hbar} S_{\alpha,\gamma_E}(\br,\br^\prime,E)
-i\frac{\pi}{2}\nu_{\alpha,\gamma_E}}\nonumber
\end{eqnarray}
where

\begin{equation}
g_{\alpha,\gamma_E} = \sqrt{\left|-\frac{1}{\dot x^{\alpha}_{n\alpha,\gamma_E} \dot x^{\alpha\prime}_{n\alpha,\gamma_E}}\right|}
\label{endepgf2degB}
\end{equation}
From the energy dependent Greens function may then be found the oscillatory part of the density of states:

\begin{eqnarray}
 d_{\mathrm{osc}}(E) & = & 2\int \sum_{\alpha,\gamma_E^\circ}\frac{d^dx dp_{1,\alpha,\gamma_E^\circ}...dp_{n-1,\alpha,\gamma_E^\circ}}
{(2\hbar\pi)^d}\frac{1}{|\dot x_n^{\alpha,\gamma_E^\circ}|}\nonumber\\
&\times&
 \mathrm{Im}\left(ie^{\frac{i}{\hbar}S_{\alpha,\gamma_E^\circ}
-i\frac{\pi}{2}\nu_{\alpha,\gamma_E^\circ}}\mathrm{tr}(\hat h_{\alpha,\gamma_E^\circ})\right).
\label{densityofstatesdegenerate}
\end{eqnarray}
For problems with only one non-cyclic coordinate the 0 length orbit contribution to the density of states has a very similar form

\begin{eqnarray}
 d_0(E) & = & \int \frac{d^dxd^dp_\alpha}{(2\pi\hbar)^d}\delta(E-H_\alpha)\label{d0}
\\
 & = &\sum_{\gamma_E^\circ}\int\frac{d^dxd^dp_\alpha}{(2\pi\hbar)^d}
 \delta\Bigg(\left.\frac{\partial H_\alpha}{\partial p_n}\right|_{\gamma_E^\circ}(p_n-p_{n\gamma_E^\circ}) \nonumber \\
 & + & \left.\frac{\partial H_\alpha}{\partial x_n}\right|_{\gamma_E^\circ}(x_n-x_{n,\gamma_E^\circ})\Bigg)\nonumber\\
 & \approx & \sum_{\gamma_E^\circ}\int \frac{d^dx dp_{1,\alpha,\gamma_E^\circ}....dp_{n-1,\alpha,\gamma_E^\circ}}{(2\pi\hbar)^d}\frac{1}{\left|\dot x_{n,\alpha,\gamma_E^\circ}\right|},\nonumber
\end{eqnarray}
where objects with the index $\gamma_E^\circ$ label different solutions to the Hamilton-Jacobi equations. Consistent with the notion that we are treating very small action orbits we have expanded the argument of the Dirac $\delta$-function about zero position and momentum, deployed a $0$ length approximation $(x_n-x_{n,\gamma_E^\circ})\approx 0$, and subsequently used Hamilton's equation to arrive at the final result.

It is straightforward to generalize the procedure to problems with arbitrary combinations of cyclic and non-cyclic coordinates with the only change for more than one non-cyclic coordinate is the reappearance of a monodromy matrix in the density of states which is, however, then restricted to the space of the non-cyclic coordinates.


\subsection{Density of states for 1D problems: a generalized Bohr-Sommerfeld quantization condition}

In the case that the action, the semi-classical phase, and the Maslov index do not depend on initial positions (and the Hamilton-Jacobi equations are non-degenerate) then $d_0(E)$ and $d_{osc}(E)$ may be straightforwardly combined and the Dirac comb identity used to yield an expression for density of states in terms of delta functions:

\begin{eqnarray}
 d(E) & = &d_0(E)+ d_{\mathrm{osc}}(E) \label{density_states}
 \\
      & = &\int \sum_{\alpha,\gamma_{E,p}^\circ,n}\frac{d^dx dp_{1,\alpha,\gamma_{E,p}^\circ}...dp_{n-1,\alpha,\gamma_{E,p}^\circ}}
{(2\hbar\pi)^d}\frac{1}{|\dot x_n^{\alpha,\gamma_{E,p}^\circ}|} \nonumber \\
      & \times & \delta\left(\frac{S_{\alpha,\gamma_{E,p}^\circ}}{2\pi\hbar}-\frac{\nu_{\alpha,\gamma_{E,p}^\circ}}{4}+\frac{1}{2\pi}\int_0^{T_{\alpha,\gamma_{E,p}^\circ}}dt M_\alpha-n\right),\nonumber
\end{eqnarray}
where $\gamma_{E,p}^\circ$ denotes once more a primitive orbit. From the $\delta$-function one can read off a generalization of the Bohr-Sommerfeld quantization condition which is given as

\begin{equation}
 \frac{S_{\alpha,\gamma_{E,p}^\circ}}{2\pi\hbar}-\frac{\nu_{\alpha,\gamma_{E,p}^\circ}}{4}+\frac{1}{2\pi}\int_0^{T_{\alpha,\gamma_{E,p}^\circ}}dt M_\alpha-n = 0
\label{BS}
\end{equation}


\subsection{Summary of semi-classical steps towards the density of states}

We briefly present a summary of the steps required to obtain the oscillatory density of states for a generic matrix Hamiltonian $\hat H(-i\hbar\partial_i,x_i)$:

\vspace{0.1cm}
\emph{Reverse quantization}: In the Hamiltonian $\hat H(-i\hbar\partial_i,x_i)$ replace $-i\hbar \partial_i\to p_i=\partial_i S$. One thus finds the matrix $H(p_i,x_i)$.

\vspace{0.1cm}
\emph{Introducing a set of classical particle types}: Determine the eigenvalues $H_\alpha$ of $H(p_i,x_i)$ and corresponding normalized eigenvectors $\hat V_\alpha$ for $E\hat V=\hat H(p_i,x_i)\hat V$. In the case of degenerate eigenvalues $H_\alpha$ orthonormalize the corresponding eigenvectors and write them next to each other as columns giving the ``full eigenvector'' $\hat V_\alpha$ (an $m_\alpha \times n$ matrix with $m_\alpha$ the degeneracy number of the $\alpha$'th set of distinct eigenvalues).

\vspace{0.1cm}
\emph{Solving the classical problems}: The eigenvalues $H_\alpha$ correspond to Hamilton-Jacobi equations $E=H_\alpha$, which must be solved for the 
actions $S_{\alpha,\gamma^\circ_E}$ of all periodic orbits $\gamma^\circ_E$ at energy $E$.

\vspace{0.1cm}
\emph{Determining the Maslov indices}: Calculate the Maslov index $\nu_{\alpha,\gamma_E^\circ}$ for each orbit, which is given by the sum of all sign changes of (i) $\dot r^\parallel_{\alpha,\gamma_E^\circ}$ (the velocity along the orbit) and (ii) $\mathrm{det}(\mathbb{1}_{2(d-1)}-\mathbb{J}_{\alpha,\gamma^\circ_E})$ (see Eq.~\eqref{denstates} for the definition of $\mathbb{J}_{\alpha,\gamma^\circ_E}$). In a the case of a one dimensional problem the Maslov index is just the number of classical ``wall-reflections'' (i.e. sign changes of $\frac{\partial p}{\partial x}$) along the orbit in phase space.

\vspace{0.1cm}
\emph{Calculating the semi-classical phase}: Express the $V_{\alpha,\gamma^\circ_E}$ in terms of $x_i$ and use Eq.~\eqref{SCPH} to calculate $M_\alpha$.

\vspace{0.1cm}
\emph{The density of states}: The expressions resulting from the previous steps must then be entered into Eq.~\eqref{denstates} or, in the case of non-isolated orbits and cyclic coordinates, into Eq.~\eqref{densityofstatesdegenerate} and the integrals over the cyclic coordinates performed.

The procedure for obtaining the semi-classical Greens functions is almost the same, however, it includes non-closed orbits and, as is well known (see for example Ref.~\onlinecite{cvitanovic}), classification of all possible such orbits is a difficult problem, and this procedure is rarely used to explicitly evaluate the Greens function.

The above steps present a systematic ``recipe'' for calculating the semi-classical density of states of an arbitrary matrix valued Hamiltonian. In the next two sections we will apply this procedure first to a number of systems for which the exact quantum mechanical result is known (Section III), as well as subsequently (in Section IV) to a problem, the one-dimension strain moir\'e in bilayer graphene, for which the quantum result may only be obtained numerically (the semi-classical result, however, remains of simple analytical form).


\section{Semi-classics for exactly solvable systems}

As a first test of the semi-classical procedure outlined in the previous sections we consider a number of cases for which the exact analytical solution is known. 

\emph{Single layer graphene}: We first consider a single layer of graphene in a uniform out-of-plane magnetic field. The Hamiltonian is thus simply the Dirac-Weyl operator with minimal substitution:

\begin{equation}
 \hat H_g=v_F\vect{\hat\Pi}\vect\sigma; \quad \Pi_1=\hbar k_1+eBx_2;\quad \Pi_2=-i\hbar\partial_2,
\label{graphene-landau}
\end{equation}
and where we have employed the Landau gauge so that $k_1$ is a good quantum number of the problem. This system has been treated by Carmier and Ullmo\cite{Carmier-ullmo}, and we thus omit details of the derivation. The final result, which agrees with that given in Ref.~\onlinecite{Carmier-ullmo}, is given by

\begin{equation}
\begin{split}
 d(E)&=\sum_{n=-\infty}^\infty\frac{A|E|}{\hbar^2v_F^2}\delta\left(\frac{E^2}{2v_F^2eB\hbar}-n\right)\\
&=\frac{eBA}{2\pi\hbar}\sum_{n=-\infty}^\infty\delta(E-E_n)
\end{split}.
\label{graphenedensstates2}
\end{equation}
\emph{4-band model of bilayer graphene}: A much more difficult system to treat in any method that employs a matching procedure is the full four-band model of AB stacked bilayer graphene (in Ref.~\onlinecite{Carmier-ullmo} only the two band down-folded version of the full Hamiltonian was treated). We take the simplest model of this material in which the interlayer coupling matrix $T$ is independent of momentum (although lifting this condition would not significantly complicate the analysis)

\begin{equation}
\hat H_{\mathrm{ABbi}}=\begin{pmatrix}
\hat H_g&T\\T^\dagger&\hat H_g
\end{pmatrix};\quad T=\begin{pmatrix}0&\tau\\0&0\end{pmatrix},
\label{ham}
\end{equation}
where $\tau$ describes the interlayer hopping. With the method outlined in the previous section, this system yields straightforwardly to a semi-classical analysis as we now show. We first send $-i\hbar \partial_i\to p_i=\partial_i S$ and diagonalize the resulting Hamiltonian to find the eigenvalues

\begin{equation}
 E = \sigma_1\left(\frac{t}{2}\right) + \sigma_2\left[\left(\frac{t}{2}\right)^2 + (v_F \Pi)^2\right]^{(1/2)},
 \label{BE}
\end{equation}
where $v_F \Pi = v_F\left[(\hbar k_1 + e B x_2)^2 + p_2^2\right]$ and $\sigma_{1,2} = \pm 1$ label the four bands; the low energy chiral bands have $\sigma_1\sigma_2 = -1$ and the high energy bonding and anti-bonding bands $\sigma_1\sigma_2 = +1$. The corresponding eigenvectors are

\begin{equation}
 V_{\sigma_1\sigma_2} = \begin{pmatrix} E \\ v_F \Pi \sigma_2 e^{i\theta} \\ \sigma_1\sigma_2 E \\ \sigma_1 v_F \Pi e^{-i\theta} \end{pmatrix}
\end{equation}
where $\theta = \tan^{-1} \Pi_1/p_2$ where $\Pi_1 = \hbar k_1 + e B x_2$. These are, of course, are formally identical to the standard eigenvalues and eigenvectors of the Bernal bilayer. Eq.~\ref{BE} may be straightforwardly solved for $p_2$ and then the action found from $S = \int p_2 dx_2$ giving

\begin{equation}
 S = \frac{\pi E(E-\sigma_1 t)}{eBv_F^2}.
\end{equation}
From the eigenvectors the semi-classical $M_\alpha$ phase is immediately found to be zero (the Berry phase is also zero in this system). Note this contrasts with the case of single layer graphene where the Berry phase is $\pi$, and the semi-classical phase equals the Berry phase, i.e. the dynamical phase is zero. This is a consequence of the $\delta$-function structure of the Berry curvature for single layer graphene. As the Maslov phase is evidently simply 2, the number of turning points, and as none of these depend on the initial position of the classical orbit, we can deploy Eq.~\ref{BS} to immediately find the semi-classical spectrum

\begin{equation}
 \frac{\epsilon(\epsilon-\sigma_1)}{C}=n-\frac{1}{2},
\label{enlevels}
\end{equation}
where $C=\frac{2\hbar eB}{\tau^2}$ and $\epsilon=\frac{E}{\tau}$. The exact quantum solution is given as

\begin{equation}
\begin{split}
 \epsilon^2&=\left[C\left(n+\frac{1}{2}\right)+\frac{1}{2}\right]\\&\pm\sqrt{\left[C\left(n+\frac{1}{2}\right)+\frac{1}{2}\right]^2-C^2n(n+1)}.\\
\end{split}
\label{exactAB_stack_levels}
\end{equation}
For large $n$, the limit in which the semi-classical approximation must hold, we find 

\begin{equation}
 \epsilon=\sigma_1\sqrt{Cn}+\sigma_2\frac{1}{2}+\sigma_1\frac{1+2C}{8\sqrt{Cn}}+\mathcal{O}(n^{-\frac{3}{2}})
\label{semclassapprox}
\end{equation}
for our approximation, while from the exact result we find

\begin{equation}
 \epsilon=\sigma_1\sqrt{Cn}+\sigma_2\frac{1}{2}+\sigma_1\frac{1+2C}{8\sqrt{Cn}}+\sigma_2\frac{C}{16n}+\mathcal{O}(n^{-\frac{3}{2}})
\label{semclassapprox2}
\end{equation}
Thus the semi-classical result agrees up to $\mathcal{O}(n^{-\frac{1}{2}})$ with the exact result, and the agreement for smaller magnetic fields is better as can be seen by the fact that the higher order terms depend on $C$, which is essentially the magnetic field.

For both of these example problems (i) the Hamilton-Jacobi equations were non-degenerate and (ii) the dynamical part of the semi-classical phase vanished (and thus the semi-classical phase was identical to the Berry phase). We now turn to a problem which is both non-degenerate and, as we shall see, one in which the dynamical phase is non-zero.

\emph{Silicene}: We consider silicene, a two-dimensional allotrope of silicon with a hexagonal honeycomb structure similar to that of graphene. Spin-orbit coupling is more important in this material than in graphene (where it can generally be neglected), and therefore in the treatment that follows the spin-orbit coupling term is included. The Hamiltonian for this system is given as\cite{silicene}

\begin{equation}
\begin{split}
 H&=v_2\left(\sigma_1\otimes\mathbb{1}_2\hat p_1+\sigma_2^*\otimes\mathbb{1}_2 \hat p_2\right)\\
&-v_1\left(\sigma_3\otimes\sigma_1\hat p_2+\sigma_3\otimes\sigma_2\hat p_1+\sigma_3\otimes\sigma_3 m\frac{v^2}{v_1}\right)
\end{split},
\label{stepham}
\end{equation}
where $\otimes$ is the tensor product for matrices and we use the basis $\{|A\rangle,|B\rangle\}\otimes\{|\uparrow\rangle,|\downarrow\rangle\}$ (see Ref.~\onlinecite{silicene} for details of this model Hamiltonian). The term proportional to the Fermi velocity, $v_1$, is just the graphene Hamiltonian with a mass term $m$, while the term proportional to $v_2$ describes spin-orbit coupling.

We will consider silicene in an uniform out-of-plane magnetic field and therefore introduce into Eq.~\eqref{stepham} the minimal substitution $\hat p_1\to \hat \Pi_1=\hat p_1+eBx_2$. Replacing momentum operators by momentum functions $\hat p_i\to p_i$ in the resulting Hamiltonian, yields the matrix $H(p_i,x_i)$, diagonalization of which results in two pairs of twice degenerate eigenvalues and thus twice degenerate Hamilton-Jacobi equations. These are given as

\begin{equation}
 \begin{split}
  E=\pm\sqrt{m^2v^4+v^2(p_1^2+p_2^2)}=:\pm h
 \end{split},
\label{classeq}
\end{equation}
and evidently describe a relativistic particle with a "speed of light" $v=\sqrt{v_1^2+v_2^2}$ and mass energy $mv^2$. The corresponding ``full eigenvectors'' are 

\begin{equation}-
 V_+=\left(
\begin{array}{cc}
 0 & -\frac{ v (i p_2+p_1)}{\sqrt{2} \sqrt{ h(h-mv^2)}} \\
 \frac{ v_2 (i p_2+p_1)}{\sqrt{2} \sqrt{h(h+mv^2) }} & \frac{i  (mv^2-h) v_1}{\sqrt{2} \sqrt{ h (h-mv^2)} v} \\
 \frac{ v_1 (p_2-i p_1)}{\sqrt{2} \sqrt{h(h+mv^2) }} & \frac{\left(mv^2-h\right) v_2}{\sqrt{2} \sqrt{ h (h-mv^2)} v} \\
 \frac{\sqrt{h (h+mv^2) }}{\sqrt{2} h} & 0 \\
\end{array}
\right) 
\label{v+}
\end{equation}
and 

\begin{equation}
 V_-=\left(
\begin{array}{cc}
 0 & -\frac{ v (i p_2+p_1)}{\sqrt{2} \sqrt{ h(h+mv^2)}} \\
 \frac{ v_2 (i p_2+p_1)}{\sqrt{2} \sqrt{h(h-mv^2) }} & \frac{i  (h+mv^2) v_1}{\sqrt{2} \sqrt{ h (h+mv^2)} v} \\
 \frac{ v_1 (p_2-i p_1)}{\sqrt{2} \sqrt{h(h-mv^2) }} & \frac{\left(h+mv^2\right) v_2}{\sqrt{2} \sqrt{ h (h+mv^2)} v} \\
 \frac{\sqrt{h (h-mv^2) }}{\sqrt{2} h} & 0 \\
\end{array}
\right).
\label{v-}
\end{equation}
The action is easily found from these Hamilton-Jacobi equations, Eq.~\eqref{classeq}, to be $S_{\alpha,r}=r\pi\frac{ E^2-m^2v^4}{2eBv^2}$, and the Maslov index to be $2r$ with $r$ the number of circuits of one primitive orbit. From the ``full eigenvectors'' we find the (matrix valued) semi-classical phase to be

\begin{equation}
\begin{split}
\int_0^Tdt M_\pm &=-i\pi\frac{\sqrt{E^2-m^2v^4}}{ E}\mathbb{1}_2\\
&=-i\pi\mathbb{1}_2-\frac{1}{2}\frac{m^2v^4}{E^2}\mathbb{1}_2+\mathcal{O}\left(\frac{m^4v^8}{E^4}\right)
\end{split}.
\label{fullorbit}
\end{equation}
This expression is diagonal (yet this is generally not the case, see, for example Ref.~\onlinecite{Keppeler-Bolte}). On the other hand the Berry phase can also be calculated directly from the ``full eigenvectors'' and is found to be

\begin{equation}
\begin{split}
\int_0^TdtM_\pm^1&=-i\pi\mathbb{1}_2-i\frac{mv^2}{E}\sigma_3
\end{split}
\label{semclassberry}
\end{equation}
which clearly does not coincide with the full semi-classical phase Eq.~\eqref{fullorbit}. It is interesting to note that both phases coincide in the limit $E\gg mv^2$, where they limit to the energy independent constant $i\pi\mathbb{1}_2$, but that the next order of the semi-classical phase is quadratic in $1/E$ while it is linear for the Berry phase.

\begin{figure}[H]
\centering
\includegraphics[width=0.5\textwidth]{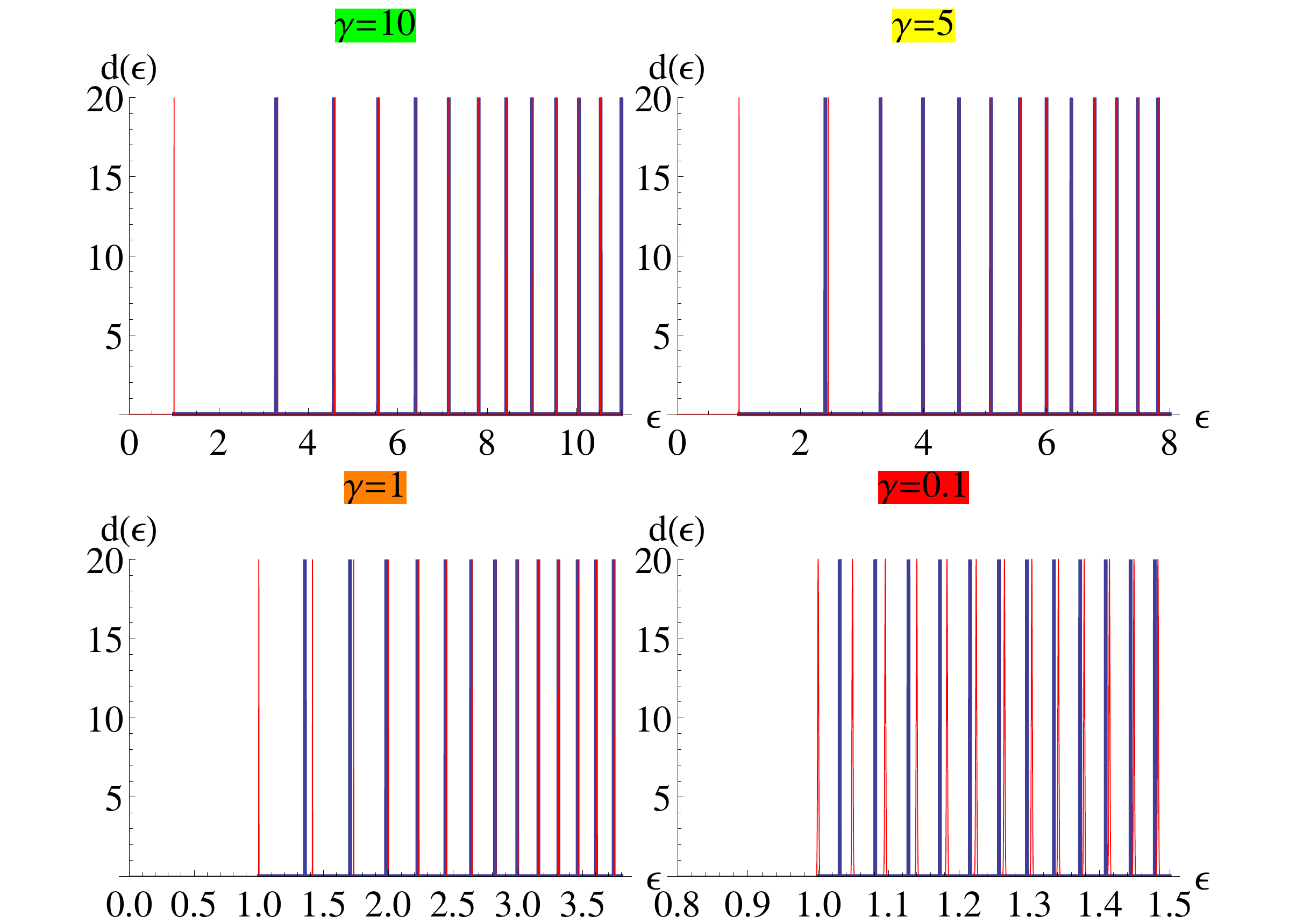} 
\caption{Plot of the exact (red) and semi-classical (blue) silicene density of states for different values of $\gamma=\frac{2eBv^2}{m^2v^4}$.}
\label{f1}
\end{figure}

As both the semi-classical phase and the Maslov index do \emph{not} depend on initial positions we may use directly the generalized Bohr-Sommerfeld quantization condition, Eq.~\eqref{BS}. To this end we require the maximum initial momentum in the cyclic direction and this provides the bounds for the cyclic momentum integral, which can straightforwardly be found from the Hamilton-Jacobi equations as $\Pi=\frac{\sqrt{E^2-m^2v^4}}{v}$. We hence encounter 
the integral $\int\frac{dp_1^\prime}{|\dot x_2|}=\int_0^{2\pi}\frac{\Pi d\theta}{v}=\frac{2\Pi \pi}{v}$ and hence the density of states is given by

\begin{equation}
\begin{split}
d(E)=&\sum_{n=-\infty}\frac{2A\sqrt{E^2-m^2v^4}}{(\hbar v)^2}\times\\
&\delta\left(\frac{E^2-m^2v^4}{2eB\hbar v^2}
-\frac{\sqrt{E^2-m^2v^4}}{2E}-\frac{1}{2}-n\right),
\end{split}
\label{silicenedensityofstates}
\end{equation}
The exact energy levels are given as $\frac{E}{mv^2}=\epsilon=\pm \sqrt{1+n\gamma}$, where $\gamma=\frac{2eBv^2}{m^2v^4}$, and we thus see that if we retain only the energy independent part of the semi-classical phase we obtain the exact result. This implies that, as the semi-classical phase is $\mathcal{O}(1/E^2)$ and the Berry phase $\mathcal{O}(1/E)$, that the extra dynamical term in the semi-classical phase is important. In Figure \ref{f1} we present a graphical comparison of the exact and semi-classical energy levels and, as may be seen, while the disagreement is pronounced at low energies, for higher energies the agreement is, as expected, very good.

\section{The strain moir\'e}

\begin{figure}[htbp]
\centering
\includegraphics[width=0.4\textwidth]{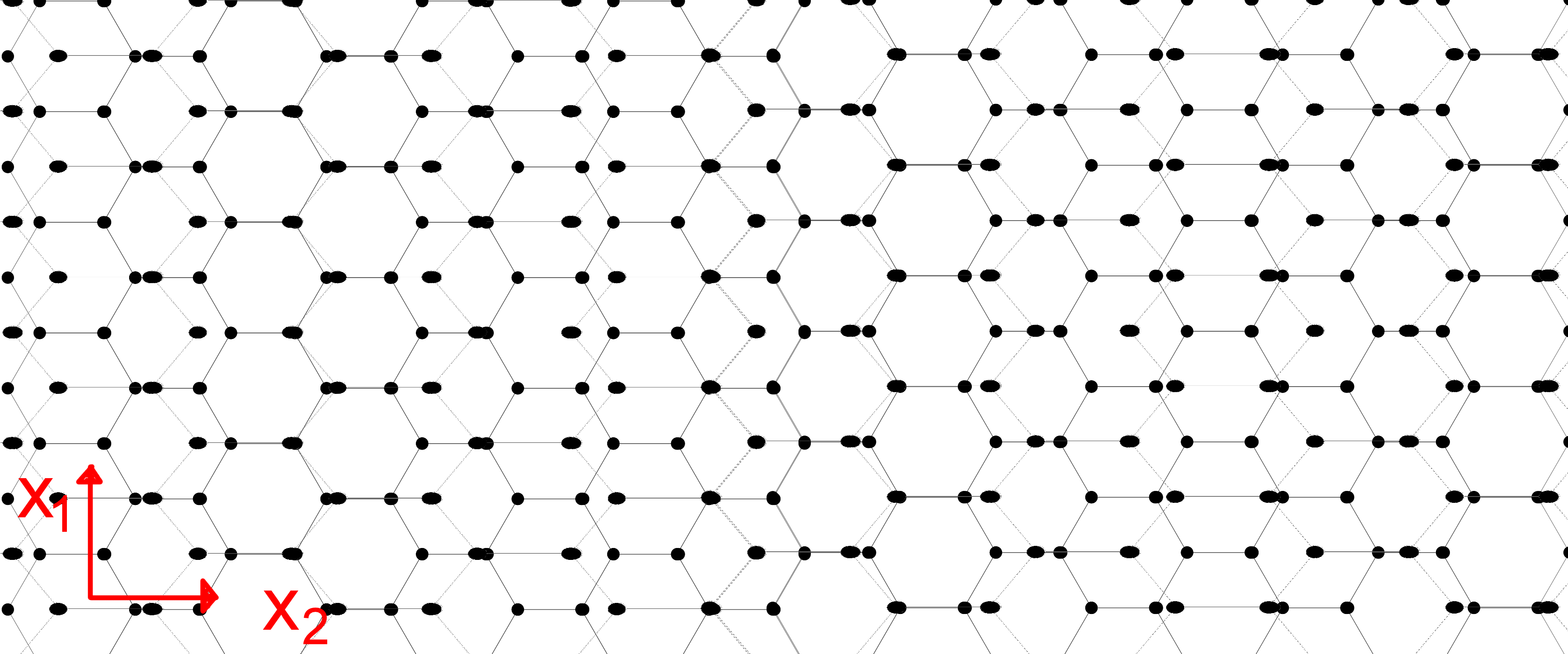} 
\caption{The graphene strain moir\'e: uniform strain is applied in the armchair direction to one layer of an initially AB stacked bilayer.}
\label{m0}
\end{figure}

\begin{figure}[htbp]
\centering
\includegraphics[width=0.45\textwidth]{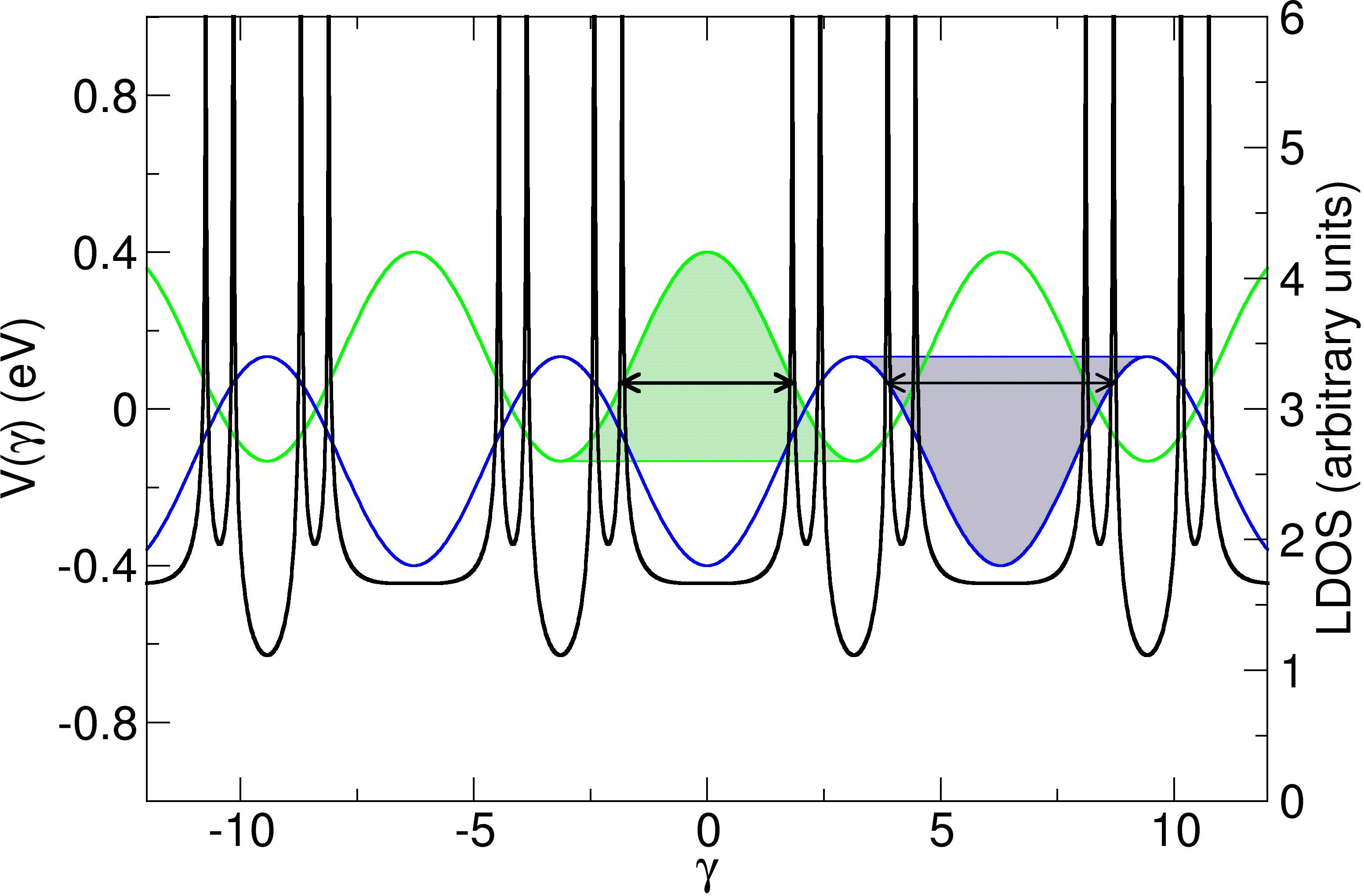} 
\caption{Structure of the classical orbits for for a strain moir\'e ($L = 1000\,$nm). The effective potentials of particle types 1 and 2 are shown as the blue and green lines respectively, with the shaded areas illustrating the classically allowed regions; note that this is inverted for particle type 2 due to the negative effective mass of that particle type, see Eq.~\eqref{mm}. For an energy of $E = 0.067\,$eV the two orbits of the two particle types are indicated, along with the corresponding local density of states. This latter quantity displays, as expected, pronounced peaks at the turning points of the classical particle. The AA spots of the moir\'e correspond to $\gamma = 2\pi n$ with $\gamma = 2\pi x_2/L$.}
\label{m2}
\end{figure}

We now consider the semi-classical analysis of a complex graphene system that is an analogue of the well studied graphene twist bilayer, the one-dimensional strain moir\'e. As may be seen in Fig.~\ref{m0} this consists of a uniform strain applied to (without loss of generality) layer 1 of an initially AB stacked bilayer that leads to a moir\'e lattice in which (as in the case of the twist bilayer) all possible stacking types occur over one moir\'e period. We choose to apply the strain in the armchair direction ($x_2$ in the coordinate system displayed in Fig.~\ref{m0}) as this makes the semi-classical analysis somewhat more tractable. The deformation field of the problem is therefore, using the coordinate system indicated in Fig.~\ref{m0}, given by $\Delta \bu(x_2)=(0,a(x_2))$, and it is convenient to express this as $a(x_2)=\sqrt{3}\frac{\gamma(x_2)}{2\pi}+\frac{1}{\sqrt3}$ with $\gamma(x_2)$ some function encoding the particular strain, and the constant shift term introduced for a more symmetrical Hamiltonian. With these definitions the interlayer potential $S(\gamma)$ can be obtained via the general theory outlined in Ref.~\onlinecite{M} with the result

\begin{equation}
S(\gamma) =
t \begin{pmatrix}
  1-\cos \gamma-\sqrt{3} \sin \gamma & 1+2 \cos \gamma \\
 1+2 \cos \gamma& 1-\cos \gamma+\sqrt{3} \sin \gamma \\
\end{pmatrix}
\label{Sfld}
\end{equation}
where $t = \tau/3$ with $\tau = 0.4\,$eV the interlayer hopping. The Hamiltonian of the strain moir\'e system is then given by

\begin{equation}
 H = \begin{pmatrix} \bsig.\bp & S(\gamma) \\ S(\gamma)^\dagger & \bsig^\ast.\bp \end{pmatrix}
 \label{HS}
\end{equation}
Note that this interlayer field is rather similar to that deployed for a the one-dimensional moir\'e treated in Ref.~\onlinecite{san12}; however in that work the moir\'e was created by shear and not by strain. Uniform strain  requires $\gamma(x_2)=2\pi\frac{x_2}{L}$, with $L$ the moir\'e period as may be deduced from Eq.~\eqref{Sfld}. In particular we have AA stacking at $2\pi n$, AB stacking at $2\pi n + 2\pi/3$, and AC stacking at $2\pi n + 4\pi/3$, as may readily be seen by substitution of these $\gamma$ values into Eq.~\eqref{Sfld}. Solving the classical $\mathcal{O}(\hbar^0)$ problem we find 4 non-degenerate Hamilton-Jacobi systems. Evidently $p_1$ is a good quantum number of the effectively one-dimensional problem, and for simplicity we will consider here the case $p_1=0$; this is not a singular limit and as such the small $p_1$ behaviour is very similar to $p_1=0$. For larger $p_1$ the system develops a much richer and interesting structure, which we will not investigate here. The 4 distinct Hamilton-Jacobi equations have 4 distinct momenta of which two differ only by a minus sign: 

\begin{equation}
 p_2^{1,2,\pm}=\pm\sqrt{2m_{1,2}^*(E)(E+V_{1,2}(\gamma))},
\label{canonicalmomenta}
\end{equation}
where the index $\pm$ indicates the sign in front of the square-root. For simplicity of discussion we will now adopt the habit of referring to the solution with index 1 and 2 as particles 1 and 2. In Eq.~\eqref{canonicalmomenta} we have expressed the momenta in terms of an effective mass $m^\ast$ given by

\begin{equation}
 m_{1,2}^*(E)=\frac{E\pm3t}{2},
 \label{mm}
\end{equation}
which is energy dependent, and an effective potential $V_{1,2}(\gamma)$ that takes the form

\begin{equation}
V_{1,2}=\pm t(1+2\cos\gamma).
\label{mV}
\end{equation}
The cosine form of this potential reveals immediately that we have a quantum well structure to the problem with the maxima (particle type 1) or minima (particle type 2) centred at the AA spots $\gamma = 2\pi n$. For particle type 1 the effective mass is negative at all $E$ where $V(\gamma)$ is defined and hence the usual regions of classically allowed and forbidden motion are inverted, and thus a maxima of the effective potential at the AA spot indicates bound orbits centred on this region of the moir\'e. Particle type 2 with a minimum of the potential well on the AA spot and a positive mass well, evidently, also describes orbits centred at the AA spot. This is illustrated in Fig.~\ref{m2} in which the shaded areas represent the regions of allowed particle motion and one can see that the situation is symmetric if we send particle type 1 to 2 and $E \to -E$ provided the sign of the mass changes as well.

To examine this situation more closely we determine the turning points of the orbits which for particle type 1 are given by

\begin{eqnarray}
 x_i^{1} & = & 2\pi n-\cos ^{-1}\left(\frac{E-t}{2 t}\right) \label{obitturningpoints1} \\
 x_f^{1} & = & 2\pi n+\cos ^{-1}\left(\frac{E-t}{2 t}\right)
\end{eqnarray}

and for particle type 2 by

\begin{eqnarray}
 x_i^{2} & = & 2\pi n-\cos ^{-1}\left(\frac{-E-t}{2 t}\right)\\
 x_f^{2} & = & 2\pi n+\cos ^{-1}\left(\frac{-E-t}{2 t}\right) \label{obitturningpoints}.
\end{eqnarray}
The orbit length for particle 2 is given by $l_{1}=2\cos ^{-1}\left(\frac{E-t}{2 t}\right)$ and (as may also be seen from Fig.~\ref{m2}) decreases with increasing $E$ until we find a zero length orbit at the band edge $E=3t=\tau$, after which the particle trajectory abruptly jumps from a zero length orbit to non-localized behavior. In contrast there are no bound states of this particle type at the other band edge of $E=-3t=-\tau$ as at $E=-t$ the bound orbits merge together, and the behavior for energies lower than this is again non-localized. For particle type 1 the situation is the same but with $E \to -E$. Interestingly, and contrary to what one might expect given the results of the previous section, the semi-classical approximation for the strain moir\'e is therefore \emph{better for lower energies than for large energies}, in particular close to band edges where the approximation is guaranteed to fail as the orbit length approaches zero. On the other hand, one should stress that for slowly varying structural perturbations - such as considered here - the semi-classical approximation is expected to be good. To calculate the semi-classical spectrum via the Gutzwiller formula we require the semi-classical phase, the actions $S$, and the orbit times $T$. We now proceed to calculate each of these in turn.

\begin{figure}[htbp]
\centering
\includegraphics[width=0.4\textwidth]{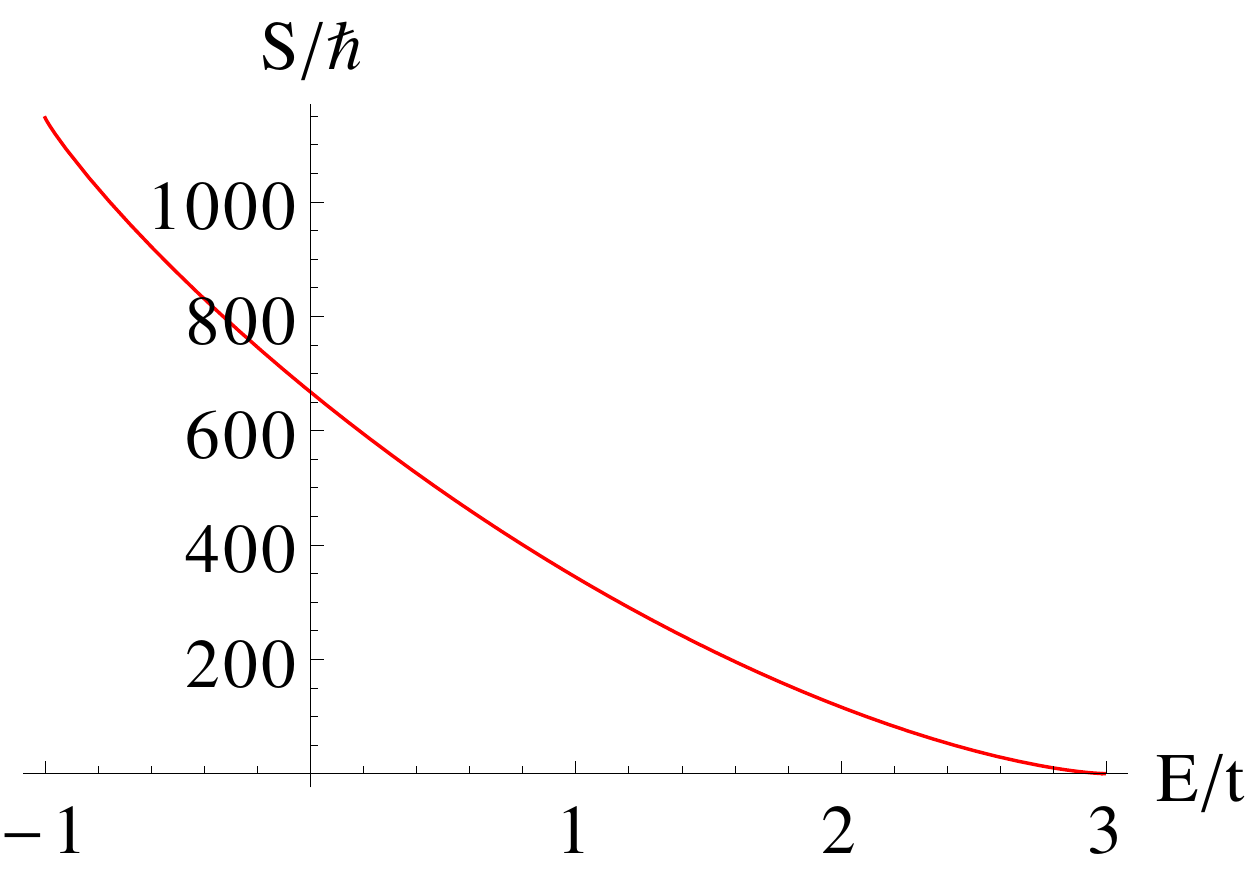} 
\caption{Plot of the action for particle type 2 in a strain moir\'e with $L=1000$nm, $v_F=0.003c$, and $\tau=0.4eV$.}
\label{Actionplot}
\end{figure}

\emph{The semi-classical phase}: The integral of the $M$ phase $I_\alpha = \oint dt M_\alpha$ may be conveniently found by the change of variables

\begin{equation}
\begin{split}
 \oint dt M_\alpha(t)&=\oint dt \hat V_\alpha^\dagger \partial_{p_\mu}H_{cl}\frac{d}{dx_2}\hat V_\alpha\\
&=\oint dt \frac{d\gamma}{dx}\hat V_\alpha^\dagger \partial_{p_\mu}H_{cl}\frac{d}{d\gamma}\hat V_\alpha\\
& =: \oint \frac{d\gamma}{\dot{x}(\gamma)} \tilde M_\alpha(\gamma)
\end{split}
 \label{Mphaseintegral}
\end{equation}
and is found to be

\begin{equation}
\begin{split}
 I_{1}= \Bigg\{\begin{matrix}-2\pi;&\quad -t<E<0\\
-\pi;&\quad E=0\\
 0 ;&\quad 3t>E>0\end{matrix}
\end{split}
\label{fitfunc}
\end{equation}
Where we also find, as we must, that $I_{2}=I_{1}(E\to-E)$. In fact only the result for $E=0$ could be obtained fully analytically; for $E > 0$ and $E < 0$ the integral was taken numerically with the values $-2\pi$ and $0$ obtained to $10^{-8}$ accuracy. 

\emph{The action $S$}: The actions may be obtained analytically from the Hamilton-Jacobi equations with the result that

\begin{equation}
\begin{split}
 S_{1}=&\frac{4 L (3 t-E) \mathcal{E}\left(\frac{1}{2} \sec ^{-1}\left(\frac{2 t}{E-t}\right),\frac{4 t}{3 t-E}\right)}{\pi  v_F},
\end{split}.
\label{actions-shorthopping}
\end{equation}
for particle type 1, where $\mathcal{E}(a,b)$ is the elliptic integral of the second kind. For particle type 2 we simply have $S_{2}= S_{1}(E\to-E)$. The action for particle type 2 is shown in Fig.~\ref{Actionplot}, and evidently is larger for smaller energies and large compared to $\hbar$, thus justifying the semi-classical approximation at low energies.

\begin{figure}[htbp]
\centering
\includegraphics[width=0.45\textwidth]{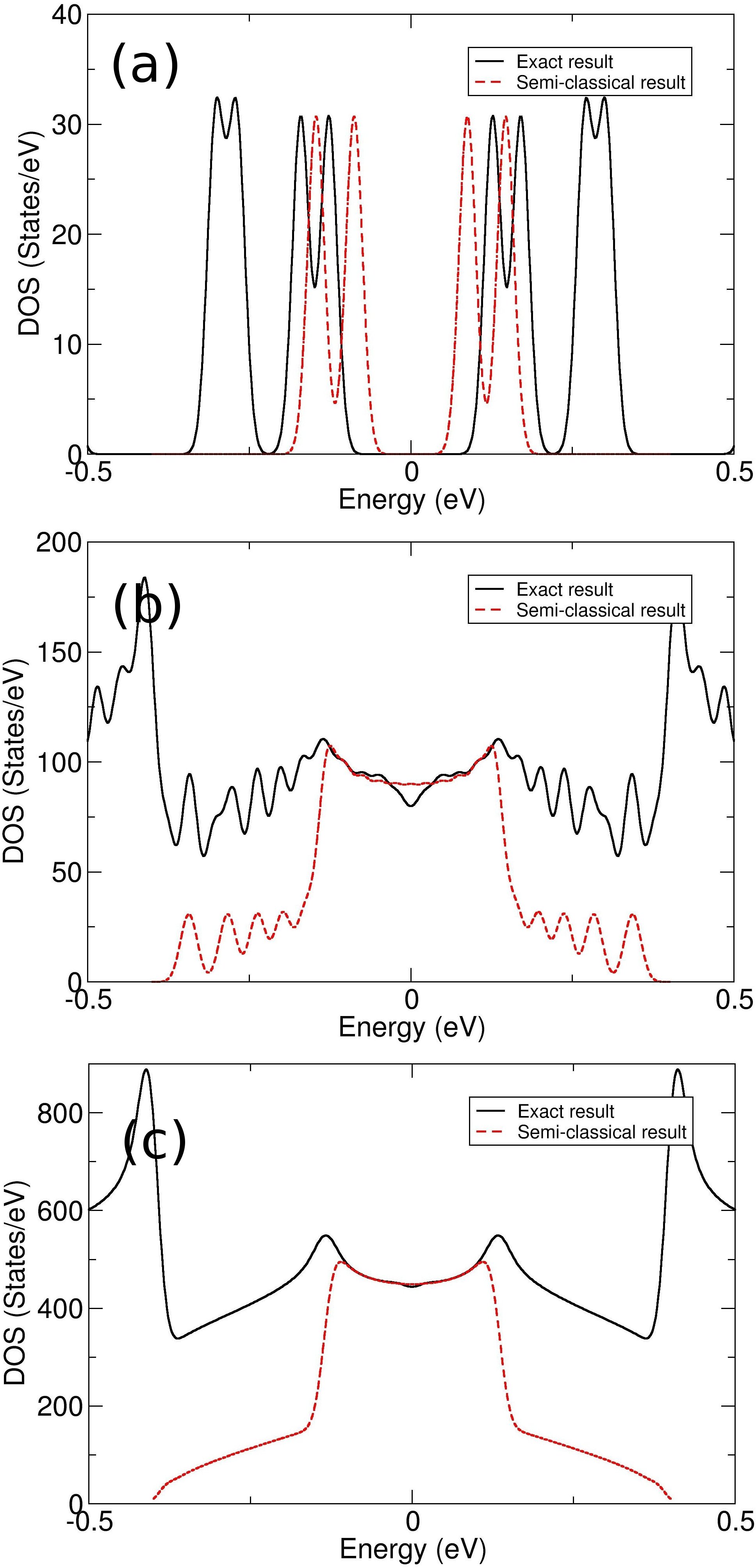} 
\caption{Plot of the Gaussian smoothed density of states for both the semi-classical approximation (red), Eq.~\eqref{densityofstatesfinal}, compared to the exact quantum result obtained by diagonalization of Eq.~\eqref{HS} (black). Shown are strain moir\'e systems with moir\'e lengths of $L=10$nm (a), $L=100$nm (b), and $L=500$nm (c) showing the increasingly good agreement at low energies between the exact result and semi-classical approximation. Note that by construction the semi-classical approximation only treats bounds orbits, and hence it is only near the Dirac point where bound orbits exist in the strain moir\'e (but at which the DOS is zero for pristine graphene) that the agreement between the semi-classical and exact solution is very good.
}
\label{m3}
\end{figure}

\emph{Orbit times}: The time for an orbit $T_\alpha$ can be found as $T_\alpha=\int dt=2\int_{x_i^\alpha}^{x_f^\alpha}dx_2\frac{1}{\dot x_2}$, where

\begin{eqnarray}
 \dot x_2^{1,2} & = & \pm\frac{v_F \sqrt{(E-3 t) (E-2 t \cos \gamma-t)}}{E-t \cos \gamma-2 t} \\
 \dot x_2^{3,4} & = & \dot x_2^{1,2}(E\to-E),
\label{velocity}
\end{eqnarray}
which follows from the Hamilton's equations and which we calculated using the Hellman-Feynman theorem
$\partial_{p_2^\alpha} H_\alpha = \partial_{p_2} \hat V_\alpha^\dagger H(p_2,x_2)\hat V_\alpha = \hat V_\alpha^\dagger \partial_{p_2} H(p_2,x_2)\hat V_\alpha$.
We thus arrive at the following expression for the period $T_\alpha$

\begin{eqnarray}
T_{1} & = & \frac{2L}{\pi v_F} \Bigg(\mathcal{F}\left(\frac{1}{2} \sec ^{-1}\left(\frac{2 t}{E-t}\right),\frac{4 t}{3 t-E}\right) \nonumber\\
          &+&\mathcal{E}\left(\frac{1}{2} \sec ^{-1}\left(\frac{2 t}{E-t}\right),\frac{4 t}{3 t-E}\right)\Bigg)\\
T_{2} & = & T_{1}(E\to-E),
\label{Time4orbit}
\end{eqnarray}
where $\mathcal{F}(a,b)$ is the elliptic integral of the first kind and $\mathcal{E}(a,b)$ the elliptic integral of the second kind.

We now have all the ingredients required to obtain the semi-classical spectrum via the Gutzwiller trace formula. As all of these quantities evidently do not depend on the initial position of the orbit we may use the generalized Bohr-Sommerfeld quantization condition, Eq.~\ref{BS}. We note that in a sample of length $L_L$ the orbits are, evidently, $L_L/L$-times degenerate and the resulting density of states is then

\begin{eqnarray}
  d(E)&=&\frac{L_L}{L}\sum_{n=-\infty}^{\infty}\left(\frac{T_1}{\hbar \pi}\delta\Bigg(\frac{S_1}{2\pi\hbar}+
\frac{I_1^M}{2\pi}-n-\frac{1}{2}\right)\nonumber \\
&+&\frac{T_2}{\hbar \pi}\delta\left(\frac{S_2}{2\pi\hbar}+\frac{I_2^M}{2\pi}-n-\frac{1}{2}\right)\Bigg)
\label{densityofstatesfinal}
\end{eqnarray}
In Fig.~\ref{m3} we plot this semi-classical density of states against the exact result, obtained simply by solving the quantum problem of Eq.~\eqref{HS} in a basis of single layer graphene states, for three moir\'e lengths, $L=10$nm, $L=100$nm, and $L=500$nm. As may be seen as the moir\'e length increases an increasing number of bound states are trapped in the AA centred quantum wells. Interestingly, even for a rather small moir\'e length of $L=10$nm the 4 bound states that are trapped in the well are reasonably well described in the semi-classical approximation. As the moir\'e length increases the agreement between the semi-classical and exact results becomes increasingly good, with at $L=500$nm the low energy agreement between the two results almost perfect.

\begin{figure}[htbp]
\centering
\includegraphics[width=0.5\textwidth]{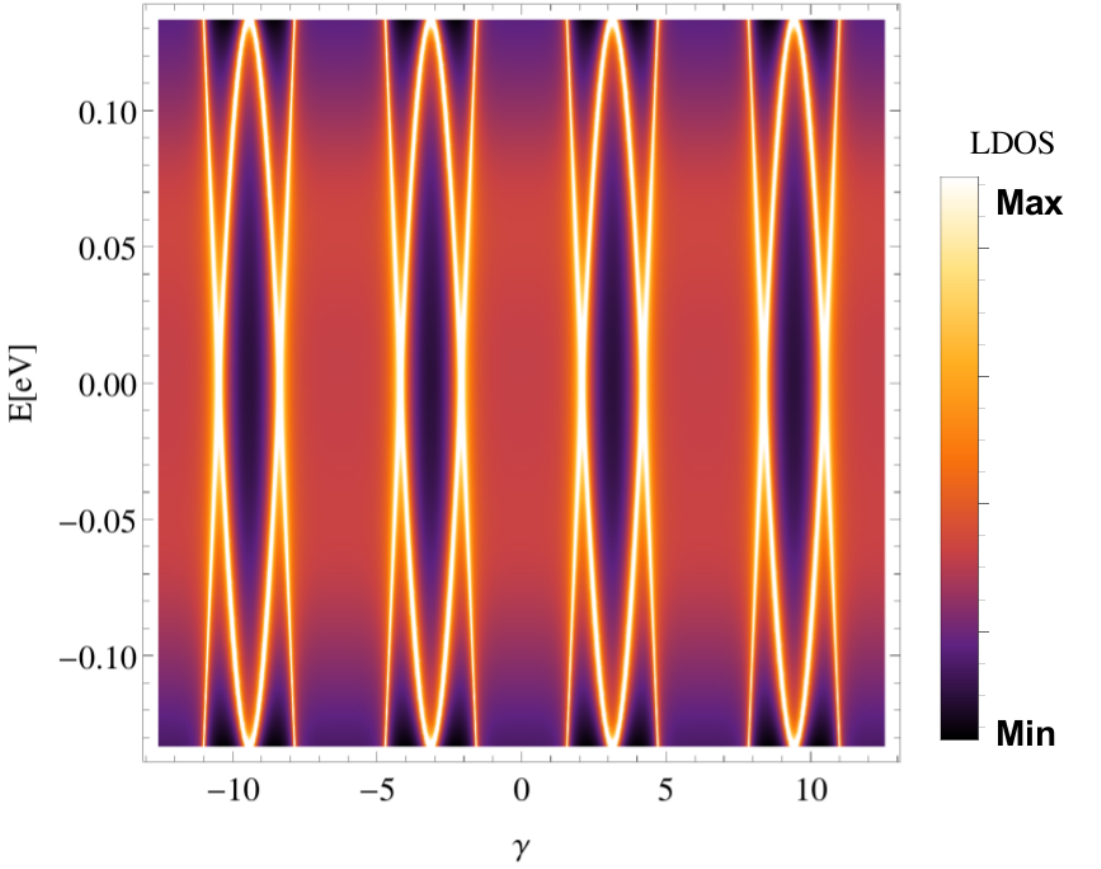} 
\caption{Plot of the smoothed local density of states for a strain moir\'e with moir\'e length $L=1000nm$, plotted in the energy position plane. The high intensity regions in the LDOS mark the turning points of the classical orbits described by Eqs.~\eqref{obitturningpoints1}-\eqref{obitturningpoints}.}
\label{m33}
\end{figure}

We may also calculate the local density of states (LDOS), obtained simply by leaving out the position trace $\int dx_2\frac{1}{\dot x_2}$ in the general formula Eq.~\eqref{densityofstatesdegenerate}:

\begin{eqnarray}
   d_L(E,x_2)&=& \frac{L_L}{L}\sum_{n=-\infty}^{\infty}\Bigg(\frac{1}{\hbar \pi\dot x_1}\delta\left(\frac{S_1}{2\pi\hbar}
+\frac{I_1^M}{2\pi}-n-\frac{1}{2}\right)\nonumber\\
&+&\frac{1}{\hbar \pi\dot x_2}\delta\left(\frac{S_2}{2\pi\hbar}+\frac{I_2^M}{2\pi}-n-\frac{1}{2}\right)\Bigg).
\label{LDOS}
\end{eqnarray}
The LDOS at an energy of $E = 0.067\,$eV is shown in Fig.~\ref{m2} along with the classically allowed orbits indicated by the arrowed lines. Evidently the LDOS shows a pronounced intensity peak at the turning points of the classical orbit, exactly as one would expect. This can also be seen in a density plot of the LDOS as a function of energy and $\gamma$, shown in Fig.~\ref{m33}, where we have smoothed the result via a Gaussian convolution for ease of plotting. The bright high intensity regions of LDOS follow the turning point structure of the classical orbits and result in regions of high LDOS surrounding the AA spots of the strain moir\'e.

\section{Conclusion}

For a general $n\times n$ matrix Hamiltonian the classical dynamics is governed by $n$ Hamilton-Jacobi (HJ) equations, corresponding to $m \le n$ 'semi-classical particle types', in a Hamiltonian phase space that is endowed with a Berry curvature. As an $\mathcal{O}(\hbar^0)$ object, i.e. a classical object, this curvature is of corresponding importance to the Hamilton-Jacobi equations in the semi-classical theory of matrix Hamiltonians, and encodes anholonomy in the transport of the eigenvectors of the object $H(\hat{\bp}\to\bp,\hat{\bq}\to\bq)$ around the classical orbits.

While for some systems the semi-classical particle types have an intuitive interpretation, e.g. in the case of the Dirac equation these correspond to electrons and positrons\cite{Keppeler-Bolte}, in a more general solid state context these simply represent the emergence of the internal semi-classical structure of the underlying quasiparticles. In particular, the number of semi-classical particle types may be less than $n$, as was the case for the strain moir\'e treated in Section IV. For each particle type $\alpha$ we find at $\mathcal{O}(\hbar^1)$ a \emph{transport equation} for the wavefunction amplitude, which differs from a similar equation found in the case of a scalar Hamiltonian by the presence of a semi-classical phase $M_\alpha$. In the case where all HJ equations are distinct, the non-degenerate case, $M_\alpha$ may be expressed as the sum of a geometric (Berry) phase that encodes global features of the $\mathcal{O}(\hbar^0)$ Berry curvature, as well as a dynamical phase that arises from the motion of the semi-classical particle ``through the Berry curvature''. This latter phase is responsible for coupling the $n$ HJ equations. In the case of degeneracies amongst the HJ systems, $M_\alpha$ becomes matrix valued and the geometric phase is then the analogue of the non-Abelian Berry phase, although the theory remains formally very similar to the non-degenerate case. 

This scheme leads to expressions for the semi-classical Greens functions and the semi-classical density of states, and we have provided such expressions for both the degenerate and non-degenerate case, as well as for further more specialized situations such as the presence of cyclic coordinates. In particular for effectively one degree of freedom systems ($n-1$ cyclic coordinates) we have presented a generalization of the Bohr-Sommerfeld quantization rule for closed orbits.

We have applied this formalism to a number of low dimensional systems: graphene, Bernal stacked bilayer graphene, silicene, and a one-dimension strain moir\'e in bilayer graphene. In the latter case we find almost perfect agreement between the exact and semi-classical density of states arising from localized states that the moir\'e induces near the Dirac point, while for the former cases good agreement is observed between the exact and semi-classical Landau spectra at high energies. Using the formalism here, even the strain moir\'e yielded rather easily to a semi-classical treatment which, we stress, would be almost impossibly complex to treat with the WKB method, or indeed any approach utilizing a matching condition between wavefunctions or Green's functions\cite{Carmier-ullmo}.

The semi-classical treatment of the strain moir\'e reveals the existence of two particle types (that dominate the negative and positive energy regions) that for energies near the Dirac point are trapped in a semi-classical potential well centred on the AA stacked regions of the moir\'e lattice. It should be stressed that such physics is in an essential way semi-classical: at the quantum level a potential well in graphene will not form bound states due to the Klein tunneling effect. The existence of such ``moir\'e potential wells'' has been discussed on the basis of numerical tight-binding calculations in which localization seen on the AA regions of the graphene twist bilayer\cite{lai12,lan13,san12}, a two dimensional analogue of the one-dimensional system considered in this work, but never rigorously shown to exist. Our work shows that this ``moir\'e potential well'' concept is in fact most naturally a semi-classical concept. This illustrates the conceptual usefulness of the semi-classical approach in providing insight into the physics of these complex systems.

As much of the structural complexity low dimensional systems is due to deformations that are spatially slow on the scale of the lattice constant, for example partial dislocation networks in Bernal stacked bilayer graphene\cite{ald13,butz14,kiss15}, the formalism presented in this work may provide not only a numerically tractable scheme for such complex systems, but also one which may yield transparent access to the underlying physics of this new class of materials. In addition, the results of this paper may facilitate a treatment of a semi-classical transport in systems with internal degrees of freedom where, in analogy to the scalar case\cite{Richter}, one should be able to find semi-classical expressions for the Kubo conductivity formula.

\begin{acknowledgments}

This work was supported by the Collaborative Research Center SFB 953 of the Deutsche Forschungsgemeinschaft (DFG).

\end{acknowledgments}


\end{document}